\documentclass[letterpaper,onecolumn,nofootinbib,11pt,showkeys]{revtex4-1}

\newcommand{\dropcap}[1]{#1}

\usepackage[margin=2cm]{geometry}
\usepackage{amsmath}
\usepackage{amssymb}
\usepackage{graphicx}
\usepackage{textcomp}
\usepackage{hyperref}
\usepackage{graphics}
\usepackage{pbox}
\usepackage{graphicx}
\usepackage{xcolor} 
\usepackage{rotating}
\usepackage{wasysym}

\DeclareRobustCommand{\sref}[2]{#1~\ref{#2}}
\DeclareRobustCommand{\ssref}[1]{\sref{Section}{#1}}
\DeclareRobustCommand{\sfref}[1]{\sref{Figure}{#1}}

\DeclareRobustCommand{\sffref}[2]{Figures~\ref{#1} and \ref{#2}}

\newcommand{\reff}{$r_\text{eff}$ }
\newcommand{\Aeff}{$A_\text{eff}$ }

\begin{document}

\title{Analysis of the Frequency and Detectability of Objects Resembling Nuclear/Radiological Threats in Commercial Cargo}

\author{Brian S.~Henderson}\affiliation{Laboratory for Nuclear Security and Policy, Massachusetts Institute of 
Technology, Cambridge, MA 02139}\email{bhender1@mit.edu}

\date{\today}

\begin{abstract}
 The threat of smuggled nuclear/radiological weapons and material in commercial
containerized cargo remains a significant threat to global security more than
a decade after the enactment of laws in the United States and elsewhere mandating
interdiction efforts.  While significant progress has been made towards deploying
passive radiation detection systems in maritime ports, such systems are incapable
of detecting shielded threats or even certain scenarios in which material
is unshielded.  Research efforts towards developing systems for detecting such threats
have typically focused on the development of systems that are highly-specific to nuclear/radiological
threats and no such systems have been widely deployed.  While most existing commercially-available cargo radiography
systems are not specifically designed for this interdiction task, if items resembling nuclear/radiological threats
are sufficiently rare in cargo radiographs to limit false alarms to an acceptably low frequency, then a smuggling
interdiction scheme based on existing technology may be feasible.  This analysis characterizes
the relevant nuclear and radiological threats that may evade detection by passive monitors, and utilizes a
dataset of 122,500 stream-of-commerce cargo container images from a 6~MeV endpoint gamma radiography system
to determine the frequency at which objects of similar size and density to such threats occur in containers.  It is found that for a broad class of threats, including
assembled fission devices, gamma radiography is sufficient to flag threats in this cargo stream at false positive rates of $\lesssim$2\%.
\end{abstract}


\maketitle

\dropcap{T}he detection of smuggled nuclear and radiological materials and weapons in maritime cargo containers
remains an unsolved problem, despite focused research and political discussion on the issue that intensified
significantly after the September 11, 2001 terrorist attacks and continues today.  While cargo containers are not the only 
method that a terrorist could use to deliver the weapons or materials necessary for an attack,
smuggling via maritime cargo is considered particularly threatening due to the fact that most
containers are loaded overseas and the combination of the large size of cargo containers with the high volume
of cargo traffic ($\sim$5$\cdot10^4$ containers/day in the US alone) provide an ideal environment
for concealing contraband \cite{arsGAO,kouzesbook}.

Estimates of the immediate economic costs alone
of a nuclear explosion in a major United States port exceed \$1 trillion \cite{randecon,abtes}, while even a much
smaller-scale attack involving a radiological dispersal device (RDD, or ``dirty bomb'')
could incur costs of up to tens of billions of dollars if trade is significantly disrupted  \cite{dirtyla}.
For these reasons, the United States government continues to pursue programs with the stated goal
of inspecting 100\% of US-bound cargo containers for the presence of nuclear weapons, or the special nuclear materials (SNM)
required to make a nuclear weapon, prior to embarkation of the containers at foreign ports as mandated by the
9/11 Commission Act in 2007 \cite{911}.  Progress towards achieving this goal, however, has been hampered by
the challenges of implementing systems in foreign ports with limited cooperation and financial support
from local governments \cite{megaportsGAO,foreignGAO,portGAO},
a lack of clear and effective goals for development of technological solutions \cite{aspGAO,asp2GAO,portGAO,100GAO,arsGAO,costGAO},
and reluctance from shippers and port operators to implement inspection regimes that delay shipments
or incur additional shipping costs \cite{aapap,aapagrp,wsc}.

A number of techniques have been proposed for the detection of smuggled nuclear materials and weapons,
although only two main classes of systems have been widely deployed: passive radiation detectors and
gamma/x-radiography scanners.  Passive detectors, while capable of detecting a variety of threatening
radioactive materials, may be defeated by surrounding the smuggled object with enough material
(``shielding'') to absorb the radiation before it leaves the container.  Radiography of containers,
which is capable of producing high resolution images of containers for both manual and automated inspection,
has been widely deployed for a variety of security applications \cite{CHEN2005810}.  Generally, however, radiography alone has been considered insufficient
for the detection of nuclear and radiological threats due to a lack of specific signatures for such threats, even in dual-energy systems
that provide limited material-type identification \cite{BENDAHAN2017242}.  This has led to considerable research efforts
to develop systems that can produce specific signatures for SNM. Concepts for such systems have included the detection of prompt neutrons from
photofission \cite{pnpf-short}, nuclear resonance fluorescence \cite{NRF_bertozzi}, and a variety of others \cite{runkle2012rattling}.  Due to the
novelty, complexity, and cost of these systems none have been deployed widely.

Given the difficulty and costs of developing and deploying highly-specific nuclear/radiological threat detection systems,
it is worthwhile to examine the capabilities of the existing commercial systems (i.e., passive radiation detectors
and gamma radiography systems) to determine the extent of their capabilities so as to prioritize future effort
and spending in this area. Specifically, given that passive radiation scanning is capable of detecting
many classes of unshielded nuclear/radiological threats \cite{kouzesbook,KOUZES2008383} and that the materials
required to shield an object from
detection will be apparent in a radiograph \cite{katzx,RISA:RISA1696,Gaukler2011,caars}, the question arises as
to whether objects in cargo resembling
threats (according to their signatures in these systems) are sufficiently rare so as to permit an interdiction system relying primarily on these existing
technologies.  To address this question, this analysis characterizes the relevant nuclear and radiological smuggling
threats and examines data from radiographic images of 122,500 stream-of-commerce cargo containers to determine the frequency
of objects in the cargo which appear consistent with nuclear/radiological threats.  To this end, the radiography system was modeled so as to allow prediction of
the radiographic appearance of typical nuclear/radiological threats in terms of their effective size and density
and the entire image set was analyzed to determine the frequency of such objects.  Assuming that the cargo stream
contained none of the objects under consideration, this frequency amounts to the false alarm rate of an interdiction
system using the methods of this analysis.  It was found that, in this container stream, radiography is capable of
distinguishing a large class of relevant threats with a false positive rate of $\lesssim$2\%.

\section*{Classifying Relevant Nuclear and Radiological Threats}

This analysis utilizes the fact that nuclear materials are typically characterized by two key characteristics: high density ($\rho\gtrsim$18 g/cm$^3$)
and high atomic numbers ($Z_\text{U}=92$, $Z_\text{Pu}=94$).  Similarly, due to the necessarily high
level of radioactivity of an effective RDD, any RDD that successfully evades passive detection will
likely require significant amounts of dense, high-$Z$ shielding or a combination of low- and high-$Z$ shielding
to capture neutron radiation \cite{kouzesbook,KOUZES2008383}.  While an ideal system would detect even trace
amounts of material present in a container, the large size of containers and high-volume of cargo traffic would
likely make such a system prohibitively costly and disruptive to trade.  Thus, it is necessary to set reasonable
detection goals and to properly assess the capabilities of existing and proposed systems of reaching these goals by
comparison to data from actual stream-of-commerce containers. The two main classes of threats and their
passive detectability are discussed here to establish criteria for the analysis of the image set.

\subsection*{Nuclear Devices and Special Nuclear Materials}

The detection of an assembled, detonable nuclear warhead in a cargo container is clearly a minimum
requirement of any interdiction system, but detection of smuggling of highly enriched uranium (HEU),
weapons-grade plutonium (WGPu), or other SNM intended for incorporation into weapons is also highly desirable.
While it is generally considered that significant samples of WGPu are detectable by passive radiation monitors
due to their strong fission neutron signature (or otherwise would be surrounded by very significant shielding), HEU produces
very little passive radiation that would be likely to reach detectors \cite{fetter,kouzesbook,KOUZES2008383,dalal2010,ANSIrpm}.
Thus, it is important to consider how the assembled ``physics packages'' of weapons would appear in a radiograph (since
these represent the minimum assembly required for nuclear detonation) as well as smaller samples of SNM alone (especially in the case
of HEU).  Due to the nature
of SNM, these samples are still most likely to appear as anomalous dense regions in radiographs.  Previous
efforts have used 100~cm$^3$ ($\sim$2~kg) as a detection goal \cite{caars}, which will used as a lower-bound test case in this analysis.
Ultimately, the lower thresholds for SNM detection should be informed by knowledge of the processes
used to construct weapons, information regarding smuggling incidents \cite{cns}, and information from
datasets such as the one considered in this work.

\subsection*{Radiological Dispersion Devices and Radioactive Sources}

Regarding the detection of a radiological dispersion device (or a radioactive sample that could be used in
such a weapon), it is assumed in this work that currently deployed  radiation portal monitors are at least as functional
as the specified standards \cite{ANSIrpm}. Under this assumption, any attempt to smuggle a quantity of radioactive material
that poses and significant threat and is unlikely to be detected by passive monitors would require shielding.
Since a strong radioactive source may be quite small\footnote{For instance, the possible
RDD isotope cobalt-60 has a specific activity of 44~TBq/gram (1100 Ci/g).}, shielding scenarios on the scale of centimeters  
must be considered, although a dispersal weapon such as a dirty bomb would likely be considerably larger and thus require commensurately
more shielding.  Thus, in this analysis of radiography data, searches for RDDs are considered as searches for the shielding that would
be required to hide them from passive detection.

\section*{Occurrence of Threat-Resembling Objects in Radiographic Images of a Commercial Cargo Stream}

Given the capabilities of passive radiation detection systems already in widespread use at ports, the
problem of detecting nuclear and radiological threats reduces to detecting objects with sufficiently
low radioactivity to be evade detection in feasible passive scanning scenarios. Such a sample may have naturally low passive
radioactivity (such as HEU) or be encased in shielding to mask the signal. As noted in the previous section, a smuggler seeking to conceal a highly
radioactive ($\mathcal{O}(\text{TBq})$) object using shielding must use significant amounts of material that will
appear as large, dense regions in radiographic images.  For
threats in involving SNM, while the materials typically emit less radiation than RDD isotopes and thus may
require less shielding to avoid passive detection, a substantial quantity of SNM will also appear as a dense region
in a radiograph.  Since radiographic systems fundamentally measure the attenuation of a beam due to the material
in a container as a 2D function of position, they are naturally suited for identifying objects by these parameters.

While utilizing radiography to search for dense objects has been proposed previously as a mechanism
for searching for nuclear and radiological threats in cargo \cite{katzx,RISA:RISA1696,Gaukler2011,caars}, such a technique
is feasible only if objects of the relevant densities and sizes are sufficiently rare in radiographs so
as to not produce a large number of false alarms.  To avoid disruption of the flow of containers in ports,
cargo security schemes have typically sought to flag $\lesssim$3\% of containers as containing items consistent with threats \cite{caars,wsc}. Additional
inspection and/or intelligence information may provide more leeway, but likely at most a few percent of containers can feasibly
be flagged for further inspection.  Previous
data have provided information regarding the mean density of entire cargo containers \cite{density,density2}, but such studies
underestimate the occurrence of dense cargoes (due to the averaging of dense regions with empty regions)
and provide no information regarding the frequency of contiguous dense regions that resemble threats.  This analysis characterizes
the expected appearance of threatening nuclear/radiological objects in radiographs and examines a large
set of high-resolution radiographic images from a 6~MeV endpoint bremsstrahlung imaging system to identify the frequency
of such objects.   If such objects are sufficiently rare, then, in conjunction with passive scanning to detect unshielded threats,
radiography is likely sufficient for identifying important classes of nuclear/radiological threats.

\subsection*{Image Dataset}

The analysis was conducted on a set of 122,500 radiographs of intermodal cargo containers, approximately
evenly split between 20 foot and 40 foot containers, produced by a Rapiscan Eagle R60\textsuperscript{\textregistered}~scanner \cite{rapov,rapiscan9}
examining rail car-borne containers entering a European port.  The container stream represented a diverse array of cargo, and approximately
20\% of the containers were empty \cite{rogers}. Each image consisted of a two-dimensional array of 16-bit pixels representing the
integrated energy transmission (measured by CdWO$_4$ detectors) of a 6 MeV endpoint bremsstrahlung beam through the cargo relative to the open
beam\footnote{The dataset also included 4~MeV endpoint images for each container, which
were also analyzed but are not discussed in this work due to the lesser penetration of the lower energy beam.}.
Each pixel represented an approximately $5\times5$~mm region on the mid-plane of the container
transverse to the beam\footnote{Due to the ``fan'' shape of the beam created by the vertical collimation of the gamma rays, objects
on the near side of the container to the source will appear larger in the vertical direction than objects closer to the detectors. For this
analysis, it is assumed that any threatening objects are near the center of the container since a smuggler would not have knowledge of
the orientation in which a container is inspected and since this placement is most advantageous for shielding the object from
passive detection.}.

This image set was previously analyzed in the context of using machine learning techniques
to identify complex objects such as vehicles in cargo images \cite{rogers,6918699,6918698}.  The analysis
was conducted on images preprocessed according to the methods discussed in Section 7.2.2 of Reference
\cite{rogers}, which trimmed the images to remove blank space around the containers and corrected for
known system artifacts in the images.  Additionally, for this analysis the images were further trimmed
to remove any portion of the rail car present in the image.  Any images with resulting sizes inconsistent
with the expected container sizes or otherwise anomalous data were excluded from the analysis.  These images constituted 
$\sim$3.1\% of the available images, and are not counted in the 122,500 total.

\subsection*{Analysis}

The analysis sought to test the hypothesis that radiological and nuclear threats may be
detected in radiographic images of cargo containers by identifying them by their anomalous
size and density relative to common cargo.  To accomplish this, the radiographic system
was modeled so as to allow the simulation of images of threatening objects and to characterize
their appearance in images.  The cargo images were analyzed for the presence of contiguous
dense regions and the frequency of regions matching the parameters of the simulated threats was examined.
If the occurrence of threat-like objects in the cargo stream is sufficiently small ($\mathcal{O}(1\%)$), then radiography provides
a stronger tool for detecting nuclear/radiological threats than previously has been appreciated.
Additionally, a number of other results regarding the distribution and
density of materials in containers follow from the analysis, which provide useful data
for various studies in cargo threat detection.

\subsubsection*{Radiographic System Model}

Due to the high intensity of photons in a bremsstrahlung radiography system, the detectors
are operated in charge summing mode (in which the signal is roughly proportional to the total
energy deposited in a detector over an integration time window), rather than counting mode (in which individual photons
and their deposited energies are detected).  The fundamental quantity measured by the system for a given material
in the beam is the transmission ratio (scaled to the 16-bit dynamic range of the detectors for the
Eagle R60\textsuperscript{\textregistered} system)
\begin{equation}
\label{eq:tran}
T_\text{mat} = \left(2^{16}-1\right)\frac{Q_\text{mat}}{Q_\text{air}}, 
\end{equation}
where $Q_\text{mat}$ is the charge sum for a fixed period with the material
in the beam and $Q_\text{air}$ is the sum for an equivalent period with no
intervening material.  In practice, radiographic systems like the one described
in this study have dynamic calibration mechanisms to account for variations in the
beam and detectors to ensure proper normalization between
$Q_\text{mat}$ and $Q_\text{air}$ \cite{rapov,rapiscan9,CHEN2005810}.
The detected charge sum may be modeled as 
\begin{equation}
 Q = C\sum\limits_{j}E_j\left(\mathcal{D}\mathcal{M}\vec{b}\right)_j + N(Q),
 \label{eq:sysresp}
\end{equation}
where $C$ is a proportionality constant, $\vec{b}$ is a vector representing a histogram of the bremsstrahlung beam spectrum as a function
of photon energy (with bin energies $E_j$), $\mathcal{M}$ is the matrix representing the effect of the material
on the transmitted beam (such that $\mathcal{M}\vec{b}$ is the transmitted spectrum histogram),
$\mathcal{D}$ is the detector response matrix (such that $\mathcal{D}\mathcal{M}\vec{b}$
would be the detected spectrum histogram if the detector could be operated in photon counting mode), and $N(Q)$ is a noise term
that in general depends on $Q$.  Models for each of these
quantities for the system and a full range of materials were developed using
simulations based on the Geant4 framework \cite{geant} and photon cross section data \cite{xcom}, the details of which
are described in \ssref{sec:si_r60}.  Since this analysis examines only average behavior over many pixels, the noise term
(which may arise from counting statistics, electronics noise, etc.) is treated as a net effect on the reconstructed
value of $T_\text{mat}$ as described in \ssref{sec:si_unc}.

Rather than work with the transmission ratio $T$, it is customary in radiography to compare
cargoes of different materials to the equivalent thickness of steel that would result
in the same value of $T$.  Using the previously mentioned models, 
a look-up table was created for each elemental material to map amounts of a given material to centimeters of steel
at equivalent $T$.  This is described in more detail in \ssref{sec:si_steel}.  This conversion was used to compute
the effective object sizes in Table~\ref{tab:objs} and is utilized 
for all further discussions of imaged cargo.  For a more detailed discussion of cargo radiography,
see Reference \cite{CHEN2005810}.

\subsubsection*{Dense Object Finding}

To measure the occurrence of dense objects in cargo and their sizes, each image was analyzed
for contiguous regions exceeding given thresholds of steel equivalent thickness ranging from 11 to 30 cm-steel
equivalent.  For each
image and thickness threshold, the image was converted to a binary image of above/below threshold and
the MATLAB\textsuperscript{\textregistered} Image Processing Toolbox was used to identify all distinct
4-connected regions in the image above the threshold \cite{conn,image}. The image toolbox was used
to determine several parameters for each region including the total number of pixels in the region, the geometric centroid, and the bounding box (smallest rectangle aligned to the image
axes that contains the region).
Each distinct 4-connected region was characterized by two parameters for comparison to the expected sizes of threat-like objects: \reff and $A_\text{eff}$.
The effective radius \reff was defined to be the radius of the largest circle centered at the centroid contained within the bounding box (scaled
to the effective pixel size along the transverse mid-plane of the container).  The effective
area \Aeff was defined to be the total number of pixels above threshold in the region times the approximate cross sectional area of a single
pixel ($\sim$0.25 cm$^2$).  The parameter \reff works well for characterizing compact objects, like a small mass of SNM, while \Aeff better
represents the material composition of elongated objects.  Note that under this definition, the image of an annular object will result in 
a value of \reff comparable to the outer radius of the object but with a smaller \Aeff than a filled-in
circular image of the same radius.

\begin{figure*}[pthb]
\centering
\includegraphics[width=0.629\textwidth]{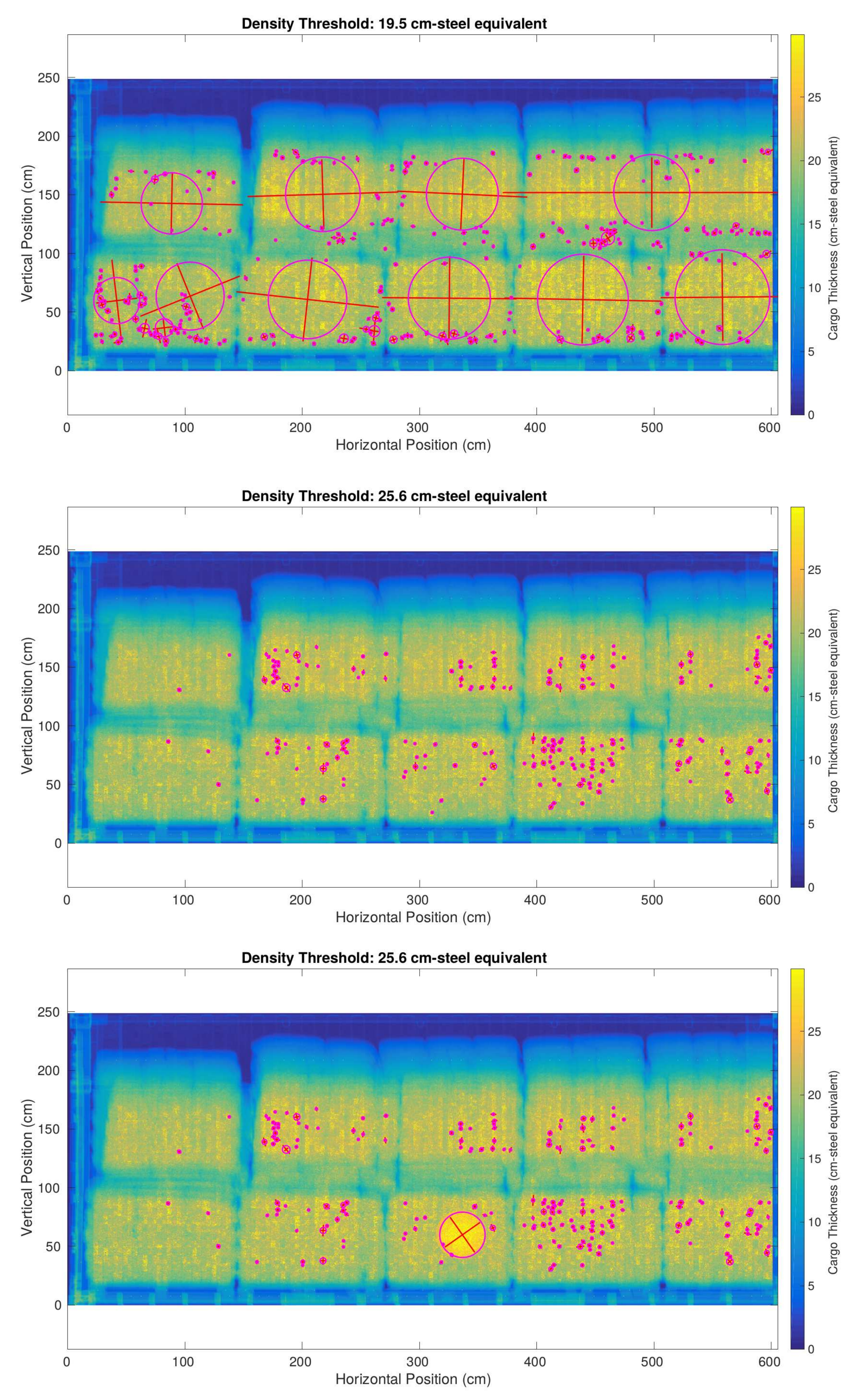}
\caption{Sample images of a 20~foot container with palletized cargo two-high stacks for which the analysis described in the text has been performed to identify contiguous dense
regions in the cargo.  In the top figure the threshold is set to 19.5 cm-steel equivalent, just below the average
thickness of the pallet stacks ($\sim$20.5 cm-steel equivalent), and thus the algorithm identifies the individual pallets as the regions above threshold.
In the middle image the threshold is set to 25.6 cm-steel,
the value used in this analysis to search for larger objects of interest such assembled fission weapons (see Table~\ref{tab:objs}).
At this higher threshold, only small ($\sim$cm-scale) regions are identified,
corresponding to signal noise and/or fluctuations in the uniformity of the cargo.  The bottom image uses the 25.6 cm threshold, but the plutonium weapon test object (\protect\sfref{fig:si_fpu}) has been
inserted into the image (accounting for saturation of the transmission, noise, etc.\ as described in \protect\ssref{sec:si_size}) and is clearly identified
by the analysis.
In each image, the red crosses mark the total extent of each
contiguous region identified by the analysis, while the magenta circle represents the circle of the effective radius $r_\text{eff}$.  Note that regions with $r_\text{eff}<2$~cm are not marked for clarity
and that not all pixels contained in the extent or circle of a given region are necessarily above threshold (the region must only be 4-connected).}
\label{fig:container}
\end{figure*}

The results of this analysis for a 20 foot container of palletized cargo at two different thresholds
are shown in Figure~\ref{fig:container}
as an example representative of a cargo type that contains significant dense regions. Additionally, the figure includes
an additional image in which a simulated fission device is inserted into the image and identified by the algorithm as 
a large dense region.  In the figure,
the magenta circles represent the circle defining $r_\text{eff}$ and the red crosses mark the
centroid and total extent of each region.

\subsection*{Characteristic Test Objects}

For concreteness in the following discussion, it is useful to define several specific examples
of potential smuggled objects.  Since a spherical object minimizes the effective radiographed area
from an arbitrary angle of a given amount of material, the example objects considered here are
each spherically symmetric.  While a smuggler could, in principle, machine material and pack a container
so as to make an object thinner along the beam direction and thus appear less dense, this would be difficult or
impossible in many scenarios.  For an example, the geometry of an assembled weapon is fixed and changing the
configuration of smuggled materials would increase the amount of shielding required to mask any passive radiation
(making the object appear larger in the radiograph).

The considered objects, summarized in Table~\ref{tab:objs}, were chosen so as to broadly represent three classes of objects: bare masses
of SNM, assembled fission packages, and Pb shields that could be used to mask
a small sample of an RDD isotope or HEU.  In the first category, near-critical masses of WGPu and HEU as well
as a 100~cm$^3$ mass of material were considered as scenarios in which material is shipped for incorporation
into a weapon after delivery.  The critical masses represent the scenario in which enough material is shipped
at once (in theory) to create a weapon, while the 100~cm$^3$ represents a smaller sample that has been the detection
goal of past development efforts \cite{caars}. The second category
models the fission packages of the nuclear weapon models described in Fetter, et al.\ \cite{fetter} as prototypical
representations of HEU and WGPu weapons.  Note that these devices represent only the essential components of a fission
device (i.e., SNM, tamper, neutron reflector, and chemical explosives) and thus a real assembled device would likely
have additional components that would increase the apparent size of the object in a radiograph.  The third category
examines shielding scenarios that could mask small samples of either RDD isotopes or uranium up to a few inches of
Pb encasing a sphere of radius 8.5~cm (so as to fit the HEU critical sphere).  Note that any scenarios involving
more shielding or shielding of larger objects would be necessarily easier to detect via these criteria.

Applying the radiographic system model, the transmission of the beam through each of the test objects was simulated
so as to produce simulated radiographic images of each object.  As described in \ssref{sec:si_size}, the simulated images
were produced with the same pixel size as the data (assuming placement in the middle of the container). A conservative
estimate of the pixel-to-pixel steel equivalent thickness reconstruction uncertainty as well as a model of the systematic
reconstruction variation due to photon source fluctuations were applied to the simulated images so as
to match the characteristics of the data images.  The same dense object finding routine used for the container images,
described in the previous section, was then applied to each test object image to determine the expected \reff and \Aeff
of each object at practical thresholds chosen for each object.
Each of these objects was simulated and analyzed under the assumption that no additional material was present in
the simulated radiographs, and thus the quoted sizes Table~\ref{tab:objs} are minimum estimates since
the walls of the container and any other cargo present would increase the apparent size of the dense
region associated with the test object in a radiograph.

\begin{table}[th]
\caption{Summary of the shielded and unshielded nuclear/radiological threats
chosen as typical examples for this analysis and their expected radiographed sizes at the highest practical
density threshold for each object.  The first three
objects are spherical masses of SNM, the latter two of which are just below a critical mass \cite{crit}.
The second two are the ``physics packages'' of the prototypical nuclear weapon models described in Fetter,
et al.\ \cite{fetter}. Each Pb shielding shell, the third set of objects, has an inner radius just large
enough to contain the HEU critical mass (although no additional material inside).  The last item combines
the 3~cm Pb shell with the HEU critical mass.}
\center
\setlength\tabcolsep{4.5pt}
\begin{tabular}{lrrrr}
\hline
Object & Outer Radius & Threshold $S$ & $r_\text{eff}$  & $A_\text{eff}$ \\
 & (cm) & (cm-steel) & (cm) &  (cm$^2$)\\
\hline
100~cm$^3$ SNM & 2.9 & 15.6 & 1.5 & 4 \\
WGPu $M_\text{crit}$ & 4.5 & 21.2 & 2.1 & 10 \\
HEU $M_\text{crit}$ & 8.5 & 25.6 & 7.5 & 166 \\
\hline
WGPu Package  & 21.0 & 25.6 & 9.6 & 272 \\
HEU Package  & 23.0 & 25.6 & 11.7 & 413 \\
\hline
3~cm Pb shell & 11.5 & 21.2 & 9.8 & 69 \\ 
6~cm Pb shell & 14.5 & 25.6 & 12.6 & 321 \\
9~cm Pb shell & 17.5 & 25.6 & 16.0 & 765 \\
\hline
HEU $M_\text{crit}$ + 3 cm Pb & 11.5 & 25.6 & 8.9 & 235 \\
\hline
\end{tabular}
\label{tab:objs}
\end{table}

\subsection*{Key Results} 
\label{sec:res}

A number of interesting results regarding the
detection of nuclear and radiological threats in commercial cargo
arise from this analysis.  Several key distributions from the image set are presented here
for discussion in the next section, while \ssref{sec:si_add} presents
further results that may be of interest to readers seeking more information
regarding the material distribution inside cargo containers.  For the purpose
of this discussion, it is assumed that no objects that would be classified
as nuclear/radiological threats were present in the analyzed images, and thus
any object resembling a threat constitutes a false positive.

\begin{figure}[ht]
\centering
\includegraphics[width=\columnwidth]{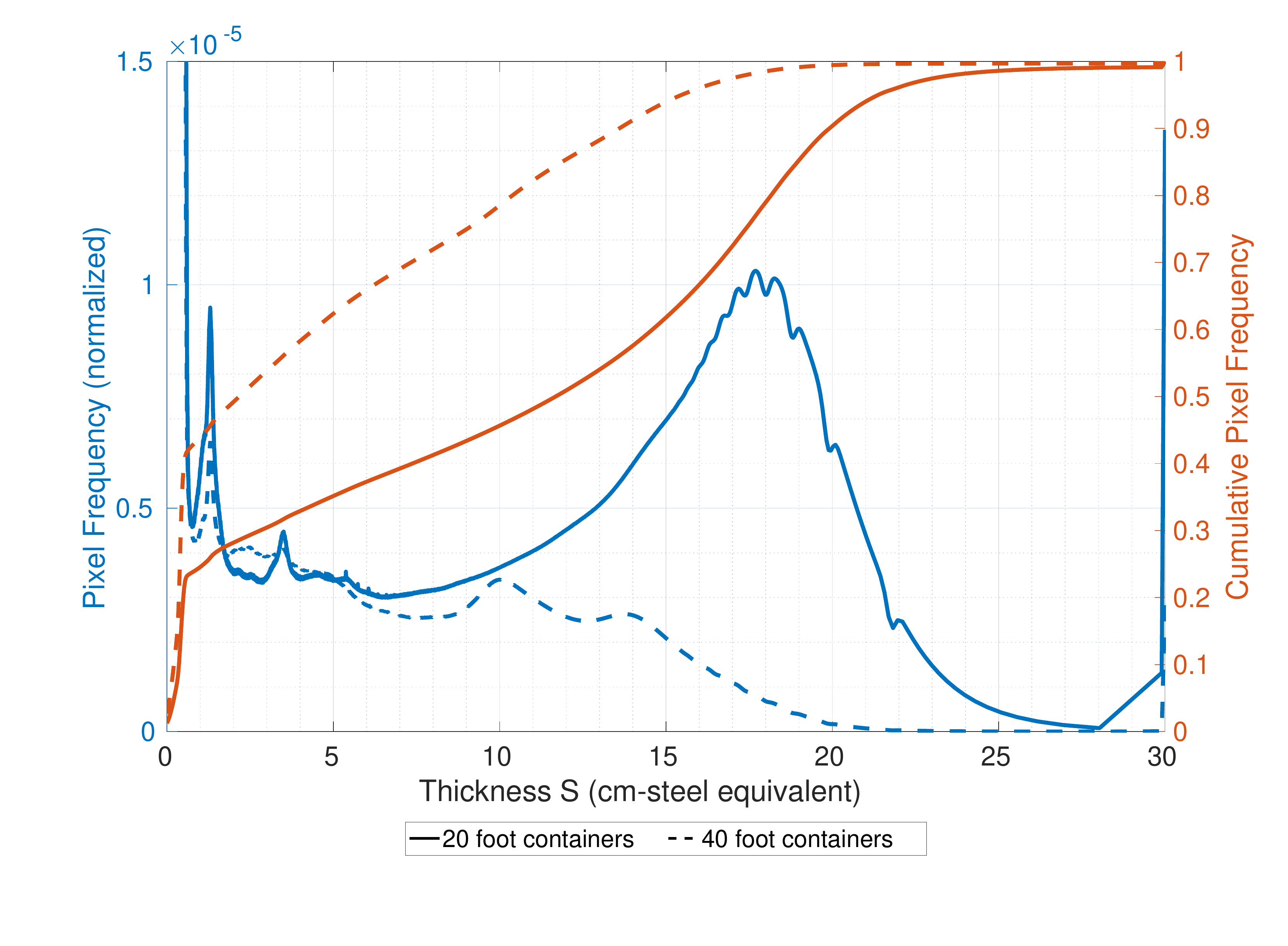}
\caption{Distributions of the effective areal density by pixel of the cargo,
separated by 20 and 40~foot containers (left axis) with the corresponding
cumulative distributions (right axis) in units of centimeters of steel
equivalent. The peak corresponding to pixels of $\lesssim$0.5
cm-steel equivalent (image pixels in which no material was present between the container
walls) is truncated for clarity; an untruncated version is shown in \protect\sfref{fig:si_upix}.  For these distributions, the portions of the container images
including the container roofs were excluded.}
\label{fig:pixdist}
\end{figure}

Figure~\ref{fig:pixdist} shows the distribution of the effective areal
density by pixel of all containers in the image set, separated by
20 and 40~foot containers, along with the corresponding cumulative distributions.
These distributions provide significantly greater detail than previously
presented cargo density distributions (such as Figure 6 of Reference \cite{density}), which
only provided average information over entire containers and thus significantly underestimated
the occurrence of high areal density regions in containers that are relevant to identifying nuclear and radiological
threats.  Notably, a very small fraction of pixels ($\mathcal{O}(10^{-5})$) exceed the quoted penetration
depth of the scanner ($\sim$30 cm-steel equivalent), and thus there are very few contiguous regions
of dense material that appear similar to the test objects.  Additionally,
40 foot containers exhibit many fewer dense pixels than 20 foot containers, which follows naturally from
the fact that 40 foot containers have twice the volume of 20 foot units but only $\sim$10\% higher payload
capacity by weight \cite{im}.

\begin{figure}[ht]
\centering
\includegraphics[width=\columnwidth]{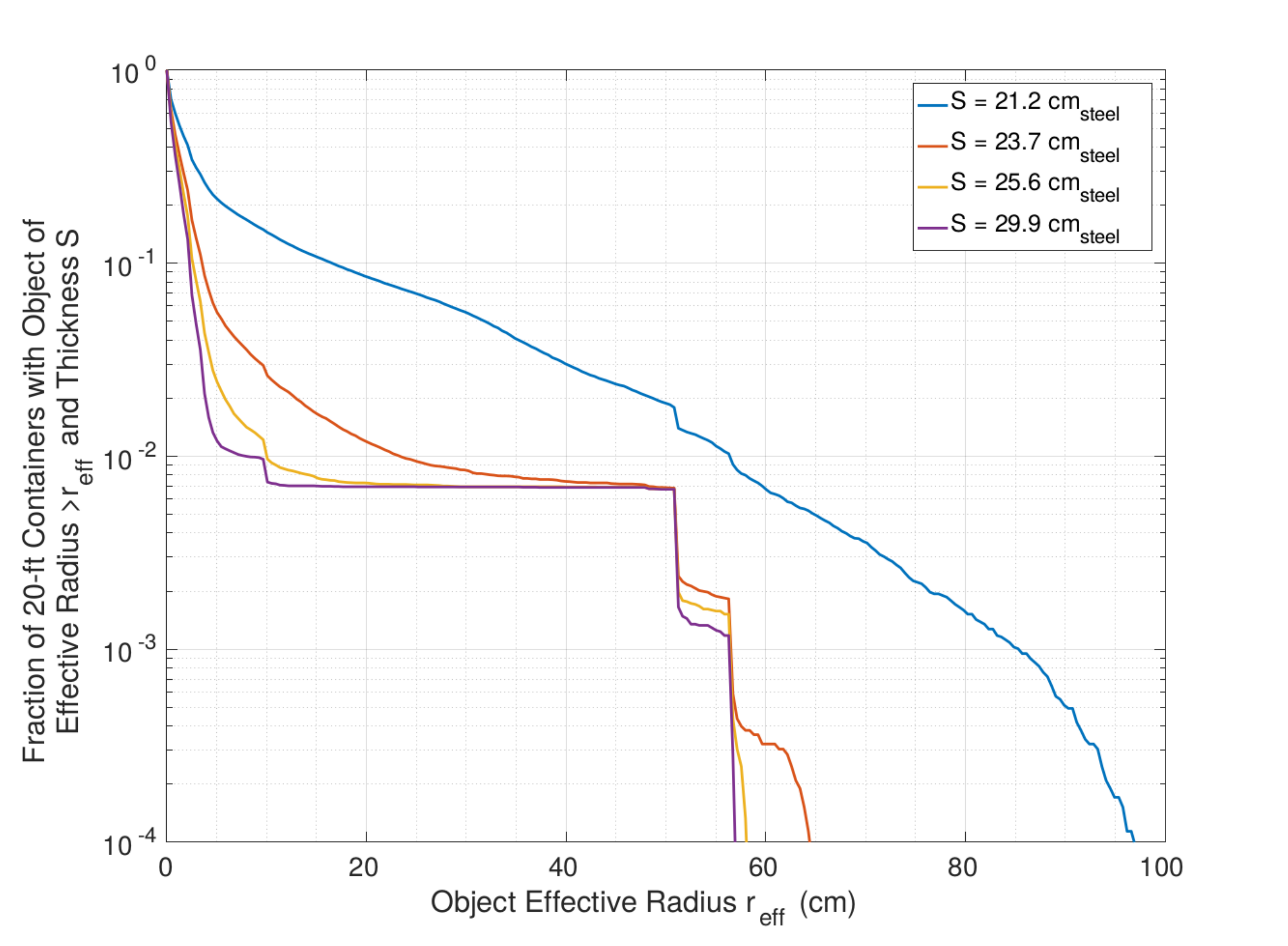}
\caption{Fraction of 20~foot container images containing a contiguous region with effective areal density ${\geq}S$ of effective radius ${\geq}r_\text{eff}$, for
several values of $S$.  See \protect\sfref{fig:r40dist} for the equivalent figure for 40~foot containers.}
\label{fig:r20dist}
\end{figure}

\begin{figure}[ht] 
\centering
\includegraphics[width=\columnwidth]{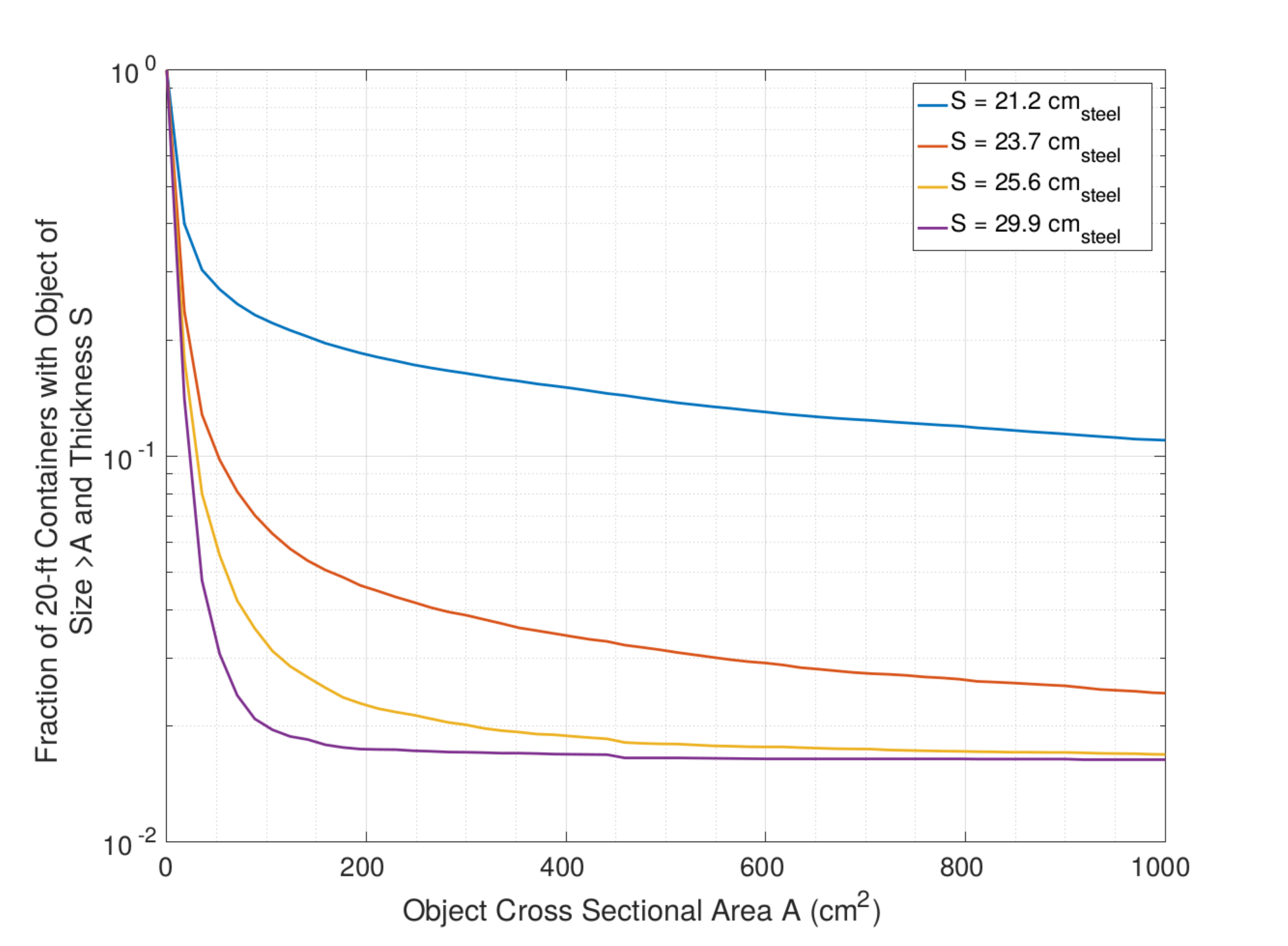}
\caption{Fraction of 20~foot container images containing a contiguous region with effective areal density ${\geq}S$ of cross sectional area ${\geq}A_\text{eff}$, for
several values of $S$.  See \protect\sfref{fig:a40dist} for the equivalent figure for 40~foot containers.}
\label{fig:a20dist}
\end{figure}

To quantify the fraction of containers that exhibit a region consistent with the test objects,
the largest contiguous region by each of the parameters $r_\text{eff}$ and
$A_\text{eff}$ at each tested density threshold was identified for each image (since a container containing at least one object consistent
with a threat would be flagged for further inspection).  The fraction of 20 foot containers whose images exhibit a
contiguous region above given density thresholds $S$ larger than \reff as a function of \reff is presented in Figure~\ref{fig:r20dist},
while Figure~\ref{fig:a20dist} presents the same for the \Aeff parameter\footnote{\label{fn:40}The same results for 40 foot containers
are presented in \sffref{fig:r40dist}{fig:a40dist}.  Across the entire analysis, 40 foot containers contain significantly fewer
dense objects of all sizes compared to 20 foot containers.}.
The fractions of containers containing an object resembling the tests objects (at the selected density thresholds) may be immediately
derived from these distributions, which amount to the false alarm rates for detection of those objects under an inspection
scheme following this analysis. Table~\ref{tab:res} presents these results.

Several results are immediately apparent in the Table~\ref{tab:res} results.  Each of the assembled fission package devices are detectable
by their effective size in either \reff or \Aeff at false alarm rates of $\lesssim$2\%, as is the spherical near-critical mass of HEU.  These
represent some of the most critical detection scenarios since detection of an assembled nuclear device would necessarily be a minimum requirement
of any effective inspection scheme, while the critical mass of HEU roughly represents an amount of material needed for a weapon and generally is
not detectable using passive radiation monitoring in even unshielded scenarios \cite{kouzesbook}.  The critical mass of WGPu presents
an unacceptably high false alarm rate when searched for using this method, but due to its strong passive neutron signature would be far more likely
to be detected using existing passive monitoring if unshielded \cite{KOUZES2008383}.  The addition of shielding or surrounding benign cargo to any of these scenarios further
reduces the false alarm rate due to the additional material increasing the apparent density and object size, making it highly likely that a highly radioactive
RDD could also be detected by either its passive radiation signature
or the radiographic signatures of the large amount of shielding (multiple cm of Pb) that would be required to mask the passive signature.
Notably, however, the data show that this method alone cannot identify smaller samples of SNM, primarily due to the fact that a spherical 100~cm$^3$ sample
of SNM has a peak steel equivalent thickness of $\sim$17 cm and thus cannot be identified as a region of unexpectedly high density (and in fact appears
as a very typical object given the areal density distributions shown in Figure~\ref{fig:pixdist}).

Note that the alarm rates presented in Table~\ref{tab:res} are somewhat conservative in that they only examine the size of
the objects in the containers, while any implemented inspection scheme could utilize a number of extra pieces of information such as
location in the container of the identified object, shape of the object, etc.\ given sufficient data describing the cargo stream.  A straightforward example
of this is apparent in the data: each of the container fraction curves for the highest density thresholds tends to level off at a constant
value ($\sim$0.6--1.8\%) before dropping to near zero at a relatively large value of \reff or $A_\text{eff}$ (for example $\sim$0.7\% of both 20 and 40 foot containers have a largest
object of high density with $r_\text{eff}\approx 55$ cm).  Examination of the cargo images reveals that these objects are elements of the container frame
that the image preprocessing failed to remove, and thus could be automatically disregarded in a search for threats.  Such an addition to the
inspection scheme could plausibly lower the false alarm rate to below 1\% for the larger test objects, although the high frequency of smaller objects
remains well above an acceptable alarm rate.

\begin{table}[th]
\caption{Percentage of 20 foot and 40 foot container radiographs that contain an object of at least the expected
size of each of the test objects listed in Table~\ref{tab:objs} by effective radius and cross sectional area.} 
\center
\setlength\tabcolsep{3.4pt}
\begin{tabular}{lrrrrr}
\hline
Object & Threshold $S$ & 20 ft. by & 20 ft. by  & 40 ft. by & 40 ft. by \\
 & (cm-steel) & $r_\text{eff}$ (\%) & $A_\text{eff}$ (\%) & $r_\text{eff}$ (\%) & $A_\text{eff}$ (\%) \\
\hline
100~cm$^3$ SNM & 15.6 & 89.80 & 100.00 & 59.94 & 100.00 \\
WGPu $M_\text{crit}$ & 21.2 & 40.78 & 39.99 & 10.09 & 10.03 \\
HEU $M_\text{crit}$ &  25.6 & 1.49 & 2.51 & 0.87 & 1.68 \\
\hline
WGPu Package  &  25.6 & 1.22 & 2.09 & 0.85 & 1.67 \\
HEU Package  &  25.6 & 0.86 & 1.88 & 0.80 & 1.66 \\
\hline
3~cm Pb shell &  21.2 & 14.94 & 24.80 & 1.06 & 2.13 \\
6~cm Pb shell &  25.6 & 0.84 & 1.97 & 0.80 & 1.67 \\
9~cm Pb shell &  25.6 & 0.76 & 1.73 & 0.80 & 1.63 \\
\hline
HEU $M_\text{crit}$ + 3 cm Pb &  25.6 & 1.32 & 2.17 & 0.86 & 1.68 \\
\hline
\end{tabular}
\label{tab:res}
\end{table}

\section*{Implications for Cargo Security Policy and Recommended Future Work} 

This work demonstrates, via the analysis of a set of 122,500 stream-of-commerce cargo
container images, that existing high-energy radiography technology, in conjunction with existing 
passive scanning methods, provides a functional means for detecting a broad class of nuclear and radiological
threats in the analyzed container stream.  The physical properties of assembled nuclear weapons, masses of SNM $\gtrsim$1 critical mass, and shielded
radiological threats sufficiently distinguish them in radiographic images 
from the typical contents of cargo  to serve as a means of identifying possible threats with false
positive rates $\lesssim$2\%.  For smaller samples of SNM, more intensive and/or specific screening methods are likely necessary.
Since radiography systems are commercially available,
overlap strongly with other customs inspection goals (e.g., detection of narcotics and stowaways), and can
be deployed at lower financial and operational cost than specialized detection systems, these results
suggest that the current development of highly-specific systems for deterring nuclear smuggling in cargo may merit reassessment.
Nuclear threat detection research continues to primarily focus on novel systems with high specificity
that may be unnecessary if it is sufficient to identify threats by their density and size as is suggested
by the data for this cargo stream.  Significant questions remain regarding the
fraction of containers that must be radiographed to deter smugglers in this
context \cite{deterfrac,RISA:RISA817}, the threshold amounts of materials that must be detected, and the operational challenges associated with deploying more radiography systems in ports. This
analysis suggests that radiography provides a stronger means of detecting nuclear threats than has been previously appreciated.  In particular, this analysis suggests that radiography offers the capability to
detect several relevant classes of threats in the near term without further technology development.

While the previous
effort in the United States to widely deploy high energy radiography systems (CAARS\footnote{Cargo Advanced Automated Radiography System}) and to upgrade passive detection systems at ports failed
due to a number of logistical and technical reasons \cite{arsGAO,asp}, the conclusions of this work support reassessment of these programs as
the most expedient and cost-effective means of reducing the nuclear/radiological smuggling threat in cargo streams.  While approximately
5\% of containers entering the United States are radiographed (having been selected based on intelligence information regarding the shipment) \cite{cbo},
many of these radiographs are conducted using lower energy radioactive isotope systems such as VACIS\footnote{Vehicle and Cargo Inspection System} that lack sufficient penetration ($\lesssim$17 cm-steel
equivalent) to identify nuclear threats by their density \cite{ORPHAN2005723}.  This suggests that systems with penetration similar to the of the 6 MeV scanner analyzed here are
necessary, but given the extremely low frequency of cargo with density $>$30 cm-steel equivalent systems with higher energies, penetration, and radiation dose are
likely not necessary in this context.  While the CAARS program and other systems under development  have focused
on identifying nuclear materials by their high atomic number \cite{mono}, which would further increase the sensitivity of a radiography system to nuclear threats,
given the capabilities of existing radiography systems for threat detection suggested by this analysis expanded deployment of standard radiography
systems may be advisable as more advanced systems are considered.  A detection regime based on this technology would be inadequate for small samples of SNM,
such as the 100~cm$^3$ sample targeted by CAARS, and thus careful consideration is required regarding the amounts of material that inspection
schemes are required to detect.  Notably, however, given that the 100~cm$^3$ represents
a concentrated amount of material and that the simulated image included no surrounding materials (i.e., benign cargo among which the sample is placed), it
is possible that in realistic scenarios additional information may be available to further increase the detectability of such small samples.  For instance,
further analysis of this data to examine the degree to which small dense regions occur in isolation or the use of radiographic systems
with multiple imaging angles for 3D reconstruction could provide sufficient information to better identify objects resembling
small SNM samples.

A lack of clear, well-motivated inspection goals has hampered research in this area, but this data provides
critical information regarding the nature of materials in containers to inform future inspection requirements.
Fundamentally, the conclusions from this analysis apply only to the commerce stream of
the analyzed images and are subject to the assumptions of the model used to determine the system response to the test objects.
Other cargo streams may contain a higher frequency of large dense
objects (e.g., a port servicing a nearby mine producing many containers filled with dense ore), and thus may
contain too many objects resembling threats to use radiography in this fashion.  Given the encouraging results of this analysis, however, examination
of images from other cargo streams, which have not been previously made available for public and/or academic
investigation, should be pursued to determine the relevance of these results for other commerce streams.  Additionally, actual radiographed images
of possible threatening objects should be utilized to eliminate model uncertainty, although such work may require a classified setting. While this analysis
focused on identifying threats by their size and density alone, further research on data of this type would likely
lead to stronger identification algorithms.  Access to data of this type provides highly useful prior information
for the development of algorithms and methods for analysis of signals for threats, not only in the context of radiography but also
in the context of passive scanning by providing detailed information about the distribution of cargo materials between sources and
detectors \cite{dalal2010}.  Additionally,
using cargo stream data to characterize and catalog benign objects would further strengthen the capabilities of a radiography-based
system to differentiate threats. As scientific and technical knowledge regarding nuclear and radiological 
weapons becomes more widespread, and thus available to potential smugglers, the availability of data to improve detection capabilities and assess existing and developing
technologies becomes more valuable.
\begin{acknowledgments}
 The author gratefully acknowledges the Stanton Foundation's Nuclear Security Fellowship program for providing the independent funding
source that made this work possible.  The assistance of Lewis Griffin and Matthew Caldwell of University College London and Edward Morton
of Rapiscan Systems was also critical as they provided access to the image set, guidance on the analysis, and computational resources.
Additionally, the author sincerely thanks Richard Lanza, Scott Kemp, and Areg Danagoulian of MIT for their guidance
on this project and numerous helpful discussions.
\end{acknowledgments}
\bibliography{refs}

\appendix
\renewcommand{\thesection}{A\arabic{section}}
\renewcommand{\theequation}{A\arabic{equation}}
\renewcommand{\thefigure}{A\arabic{figure}}
\renewcommand{\thetable}{A\arabic{table}}
\setcounter{figure}{0}
\setcounter{table}{0}

\section{Modeling of the Eagle R60\textsuperscript{\textregistered} Scanner}
\label{sec:si_r60}

In order to convert the 16-bit transmission images in the dataset to equivalent amounts of
material, and similarly to compute the expected transmission value in the Rapiscan Eagle R60\textsuperscript{\textregistered}
scanner for objects of known materials, the essential components of the radiography system were modeled: namely
the bremsstrahlung beam spectrum, the physics and geometry of the transmission of the beam through containers
mounted on the rail cars passing through the scanner, and the detector response to the transmitted beam as delineated
by Equation~\ref{eq:sysresp}.  While not all precise details of the system were available, as some elements of the system
are proprietary, a combination of published
information, known performance specifications of the system, and common functionality between gamma radiography systems
allows a suitable approximation of the system response to different materials (to within $\sim$1 cm-steel equivalent over the full penetration range)
for the purpose of the this analysis.  In particular, because the transmission measurement is fundamentally a ratio between open-beam and material-in
measurements under effectively the same conditions, errors in the model of the beam and detector response are second-order effects compared to
the effect of the transmission physics in the materials.  Each of the key elements of the model outlined by Equations~\ref{eq:tran}
and \ref{eq:sysresp} are described in the following sections.

\subsection{Setup/Geometry}

The system geometry was modeled according the technical specifications from Rapiscan \cite{rapov}, the additional information in Figure~7.2 of Reference
\cite{rogers}, and confirmed against photographs of the deployed system.  Given the fact that the vertical collimation width must be comparable to the resolution
of the images, and is thus a few millimeters, it may be considered negligible in modeling the system. Additionally, no significant variation in the transmission
value corresponding to empty container walls was observed as a function of height in the container, and thus any effects due to the vertical ``fan'' spread of the
beam that remained after image preprocessing were also considered as negligible.  While rudimentary models of the system were created for the Geant4 \cite{geant} simulations
mentioned in the following sections, no element of the geometry was critical to the ultimate analysis.

\subsection{Bremsstrahlung Beam}

\begin{figure}[ht]
\centering
\includegraphics[width=\columnwidth]{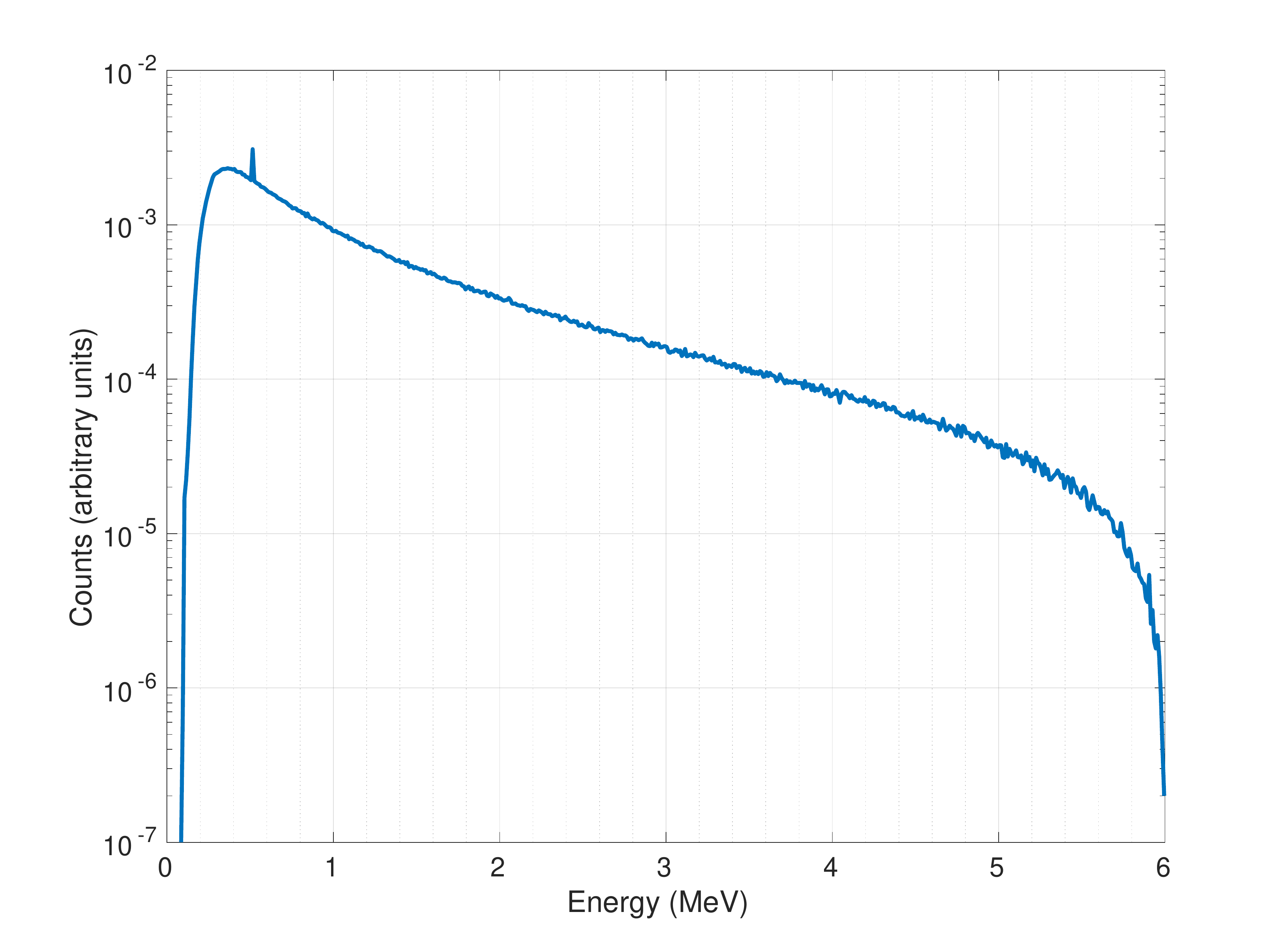}
\caption{Simulated 6 MeV endpoint bremsstrahlung spectrum used for the radiography system model.}
\label{fig:si_brem}
\end{figure}

Since the precise spectrum of the gamma rays generated by the bremsstrahlung of the Rapiscan Eagle~R60\textsuperscript{\textregistered}
system was not available, simulation was conducted to produce a suitable approximation.  A Geant4 \cite{geant} simulation was constructed
in which 6 MeV electrons (which would be produced by the linear accelerator of the radiography system) were incident on a bremsstrahlung
radiator consisting of 5~mm of tungsten backed by 5~cm of copper.  The spectrum of resulting gamma rays in a 1~degree vertical collimation
slice forward of the beam direction was recorded and is shown in \sfref{fig:si_brem}, which was used as the representation of the
beam spectrum histogram in the model ($\vec{b}$ in Equation~\ref{eq:sysresp}).

Note that since this analysis made use only of transmission ratios comparing material-in-beam to
the open beam transmission and since the images in the dataset were preprocessed to correct for
variations in the beam \cite{rogers}, only the approximate shape of the spectrum was required for
the analysis rather than its magnitude and only significant variations in the high energy portion
of the spectrum would be likely to significantly affect results for the transmission ratio.
While the absolute beam flux affects the precision of the
transmission measurement (due to variation in the number of photons detected), the uncertainty on
the reconstructed material thickness was estimated from data as described in \ssref{sec:si_unc}.

\subsection{Transmission Modeling}
\label{sec:si_trans}

The transmission component of the model ($\mathcal{M}$ in Equation~\ref{eq:sysresp}), i.e., the attenuation of the beam due to materials,
was modeled using a simple model of exponential attenuation and the NIST X-Ray Mass Attenuation Coefficients Database \cite{xcom}, which
provide photon attenuation data for all elemental materials up to $Z=92$ and energies of 20 MeV (as well as data and prescriptions for the
calculation of data for compounds of the elements). That is, for a material sample of atomic number $Z$, density $\rho$, and thickness along
the beam $X$, the attenuation of the beam at energy $E$ may be computed as
\begin{equation}
 \frac{I}{I_0} = \exp{\left(-\frac{\mu_Z}{\rho}X\right) }.
\end{equation}
For compound materials, the attenuation coefficient $\mu$ was calculated as \cite{xcom}
\begin{equation}
 \frac{\mu}{\rho} = \sum\limits_j\left( w_j\left(\frac{\mu}{\rho}\right)_j \right),
\end{equation}
where the sum runs over all elements in the compound and the weight $w_j$ is the mass fraction of the $j$\textsuperscript{th} element.  For the calculations
involving plutonium, the values of $\frac{\mu}{\rho}$ for $Z=94$ was approximated as those of $Z=92$ due to the lack of available data in the NIST
database. Using this model, for any material or mix of materials presented to the beam the matrix $\mathcal{M}$ may be constructed such that $\mathcal{M}\vec{b}$
is the spectrum after the initial beam of spectrum $\vec{b}$ passes through the material.  Note that under this exponential attenuation model, in which detection
of scattered beam components is neglected, $\mathcal{M}$ is a diagonal matrix.

This model is effective for the case of the radiography system studied here due to the tight collimation of the beam.  That is, photons reaching
the detectors are almost exclusively photons that have undergone direct transmission, and thus secondary scattering effects that are not captured
by the exponential attenuation model are not significant.  This was verified by conducting several Geant4 \cite{geant} simulations of the system for 
different materials.  The exponential transmission model differed from the Geant4 calculations by at most $\sim$0.5 cm-steel equivalent, and thus
the exponential model was used for computational efficiency across all materials.

\subsection{Detectors}

The detectors were modeled as $15.0\times 4.6\times30.0$~mm CdWO$_4$ crystals, which are typical of those used in gamma radiography systems \cite{CHEN2005810}.
In a Geant4 \cite{geant} simulation, photons of energies 0--6~MeV were impinged along the long axis of the crystal uniformly illuminating the central 5~mm of the second-longest
axis to simulate the collimated beam striking the detector crystal.  The matrix $\mathcal{D}$ mapping the initial photon energy to the distribution of resulting energy
deposition in the crystal was constructed using approximately $10^7$ simulated photons.  Like the initial bremsstrahlung spectrum, the exact details of the detectors
and their response were not available, but due to the fact that the measurement consists of a ratio and given the similarities in
detectors used across bremsstrahlung radiography systems which informed the choices for the simulation the approximation used here suffices.

\subsection{Conversion to Steel Equivalent Thickness}
\label{sec:si_steel}

With the entirety of the transmission model in place, it is straightforward to convert the transmission value $T$
calculated for any configuration of material to an equivalent thickness of steel by comparing to the calculated
transmission as a function of steel thickness at standard density ($\rho_\text{steel} \approx 8$ g/cm$^3$).  Note that
as the effective atomic number of the material of a given areal density increases, the equivalent areal density of
steel increases as well due to the increasing attenuation of photons in the beam due to increasing electron-positron pair
production cross section as a function of $Z$.  A lookup table as a function of material thickness was created for each relevant
material, which permitted direct conversion of transmission values to equivalent steel thicknesses for each of the materials.

\subsection{Uncertainty on the Steel Equivalent Measurement}
\label{sec:si_unc}

In any radiographic measurement, uncertainty on the transmission
measurement (due to the statistics associated with the number of photons
that reach the detectors and systematic effects) affects the reconstructed
effective thickness for each pixel.  Since the measured equivalent thickness
may fluctuate downward for some of the pixels representing a given object, such
fluctuations may decrease the size of the 4-connected region representing the
object.  Thus, in order to compare modeled objects to their representations
in the radiographic images, it is necessary to estimate the effects of this
reduction in the size of the 4-connected regions.

\begin{figure}[ht]
\centering
\includegraphics[width=\columnwidth]{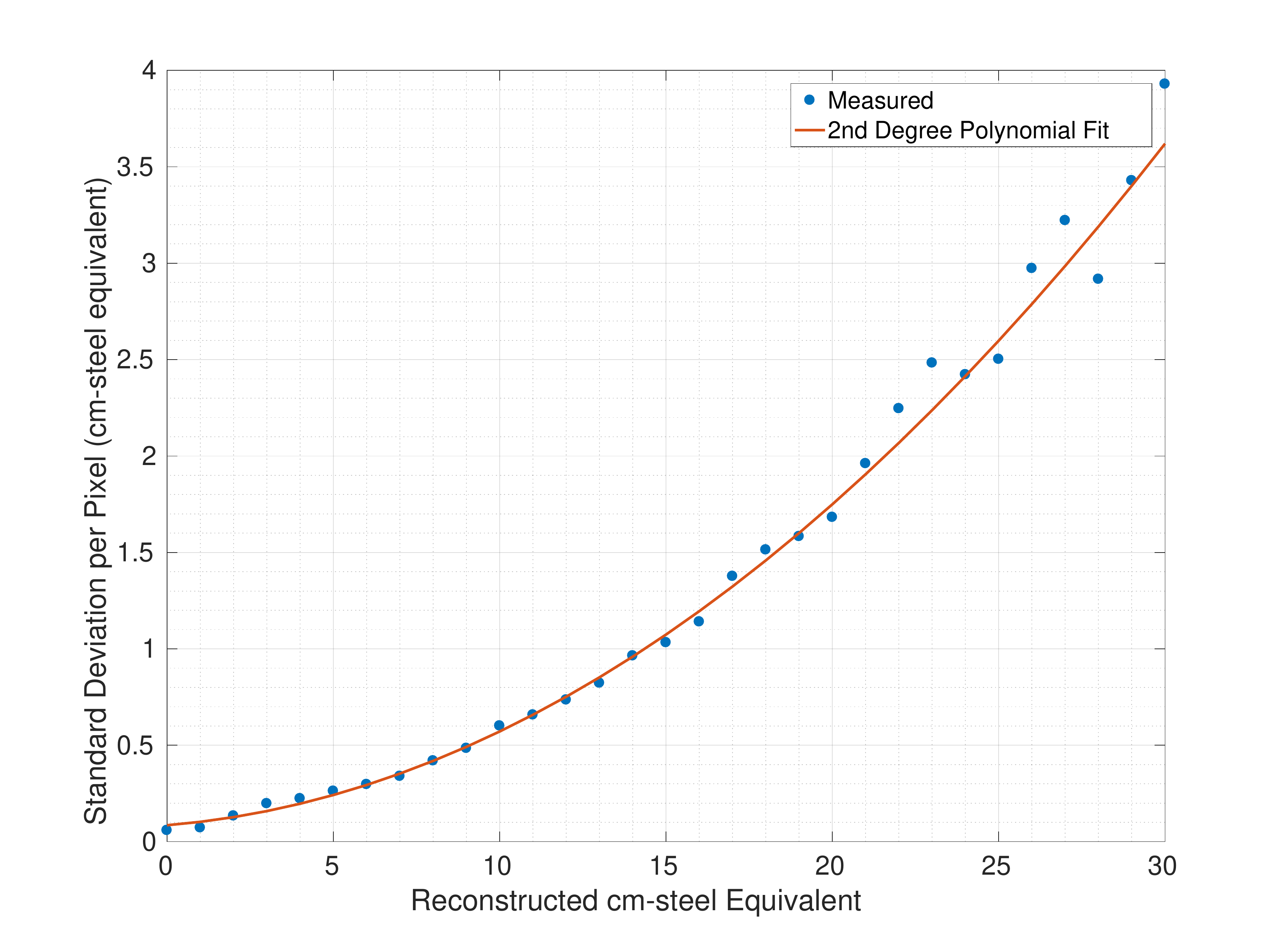}
\caption{Uncertainty on the measured steel equivalent thickness for a single pixel as a function of equivalent steel
thickness with a quadratic polynomial fit.}
\label{fig:si_unc}
\end{figure}

To estimate the uncertainty in the measurement of the equivalent steel thickness
in the images for each pixel, a data-driven approach was used to establish an
upper bound on the variance of the measurement as a function of thickness.  To
do this, each pixel in each image (for a subset of the entire image set in order
to reduce computation time) was compared to its 4-connected neighbors, and the difference
in the reconstructed equivalent thicknesses between each pixel and its neighbors was
histogrammed.  Under the assumption that adjacent 4-connected pixels represent very similar
elements of the imaged cargo given the $\sim$0.5 cm resolution of the pixels, this provides
an estimate of the measurement uncertainty but overestimates the uncertainty since adjacent pixels may in fact
represent different cargo thicknesses.  For each equivalent thickness in the range relevant to the
image set (i.e., 0--30 cm-steel equivalent in bins of 1 cm), a Gaussian was fit to the spread
of the measurements (excluding data from the $T=0$ bins) to determine an upper bound on the measurement uncertainty.
The results of this analysis are shown in \sfref{fig:si_unc}.  The combined
statistical and systematic measurement uncertainty was found to be modeled well by a quadratic polynomial
over the range of interest, as shown in the figure.  These results were applied to simulate threatening
objects in cargo as described in the following section.

\subsection{Assessing the Validity of the Model}

While this model makes a number of assumptions regarding system parameters, the model
can be compared against known features of the data and available specifications of the Eagle R60\textsuperscript{\textregistered}
system to assess its validity.  In particular, if the exponential attenuation model of the radiography system is valid, as has been
well established for narrowly-collimated bremsstrahlung radiography, then comparison to two known data points (i.e., the transmission values
for known amounts of known materials) is sufficient to test the model \cite{CHEN2005810}.  Two such points are readily available:
\begin{itemize}
 \item The largest peak in the pixel-by-pixel density distribution (which is shown without truncation in Figure~\ref{fig:si_upix})
       corresponds to an empty projection of a container in which the pixel represents the attenuation of the beam through just
       the two walls of the container and air.  Due to the highly standardized nature of cargo containers, the total thickness
       of steel present in the two walls is a known quantity.  The side walls of standard intermodal containers
       consist of corrugated steel plates that are 1.6--2.0 mm thick \cite{cspec}.  While the corrugation of the steel causes
       the amount of steel presented the beam to vary slightly along the length of the container, the average ``empty pixel''
       represents 4.3~mm of steel (due to the diagonal portions of the corrugation increasing the presented wall thickness slightly).
       The mean of the peak corresponding to the empty pixels in data occurs at a value of $T=5.59\cdot10^4$.
 \item The manufacturer of the Eagle R60\textsuperscript{\textregistered} quotes the steel penetration of the system, i.e., the greatest
       thickness of steel behind which additional material would be discernible, as 31.0~cm.  Given that measurements at the low end
       of transmission are subject to uncertainty as discussed in the previous section, this value is slightly below the value that would
       correspond to a measurement of $T=1$ (i.e., one bit out of 2$^{16}$).  From Figure~\ref{fig:si_unc}, the pixel-to-pixel measurement
       uncertainty at 31.0~cm-steel equivalent is approximately 3.8~cm-steel equivalent (standard deviation).  Thus, a measurement of $T=1$
       should ideally (under a perfect measurement) corresponds to $\sim$34.8~cm-steel.
\end{itemize}

Using the model as described, it is straightforward to compute the expected transmission values for each of these test cases.  For 4.3~mm
of steel, the model predicts $T=5.587\cdot10^4$, which is in very good agreement with the data value.  Note that due to the exponential nature
of the attenuation, the value of $T$ changes rapidly as a function of steel thickness for such small amounts of material, but even $\pm$1~mm
variations in the thickness of the walls lead to changes of $\lesssim$3\% in $T$.  Thus, the model successfully reproduces the data for the case of
the empty container walls.  At the low-transmission end of the data, the problem is different in that small changes in the value of $T$ correspond to large changes in the
amount of material.  The model predicts that 36.4~cm-steel produces a value of $T=1$, which corresponds to a slight overestimation but is within
the uncertainty of the measurement of a single pixel.  Thus, given the known data and the validity of the exponential attenuation model for the system,
the model represents the data well to within $\mathcal{O}(1\:\text{cm-steel})$ across the full range.  Due to the uncertainty at the low-transmission end of the
range, the comparisons between test objects and data in the text were conducted at density thresholds somewhat below the penetration limit of the system ($\leq$25.6~cm-steel)
to reduce any possible effect from model error in this region.

\section{Calculation of Expected Radiographed Object Sizes}
\label{sec:si_size}

\begin{figure}[ht]
\centering
\includegraphics[width=\columnwidth]{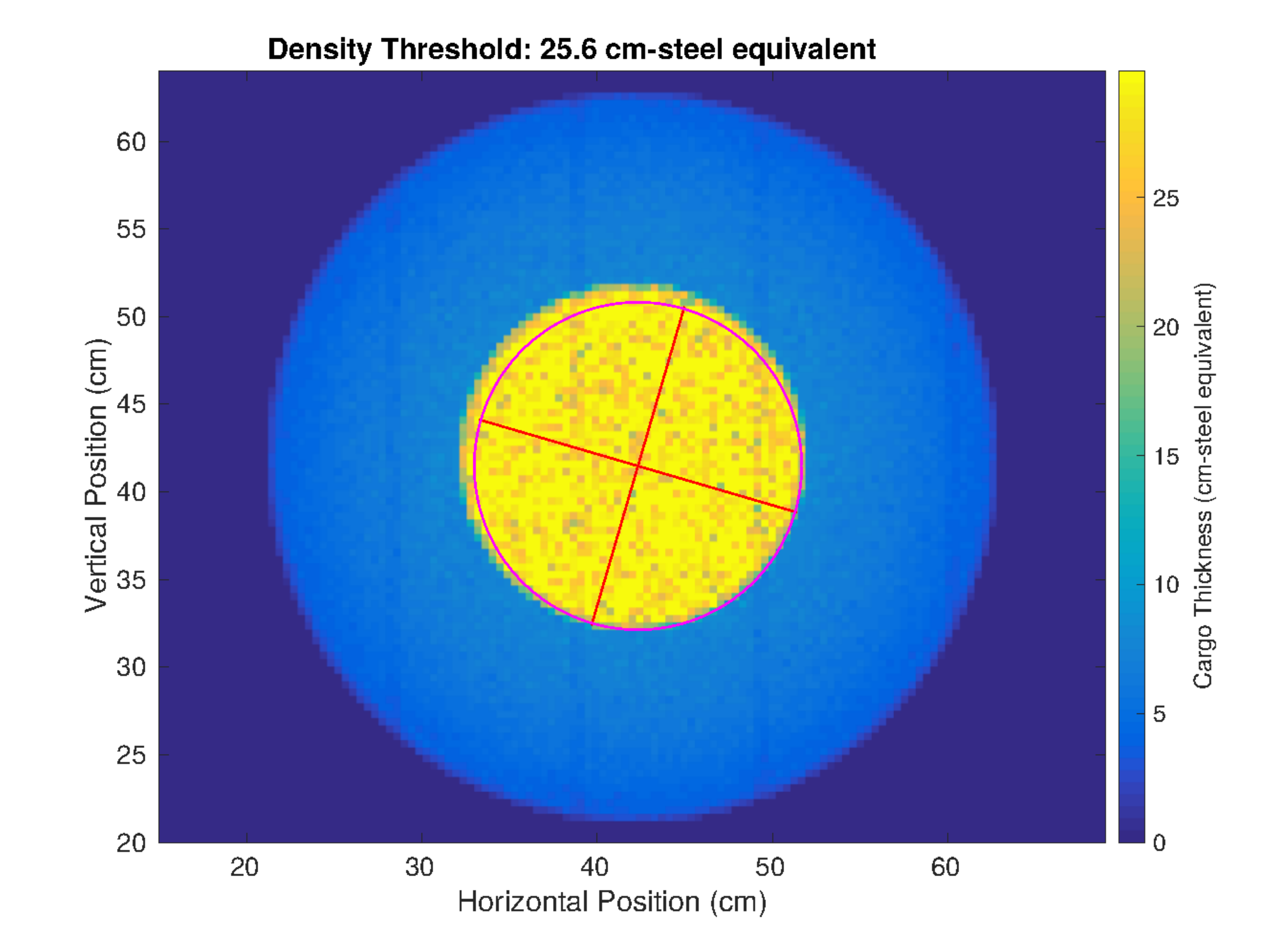}
\caption{Simulated radiograph of the weapons-grade plutonium weapon physics package
described in Reference~\cite{fetter}, with the dense object identification
algorithm applied at a threshold of 25.6 cm-steel equivalent.  The algorithm
identifies the dense central region of the weapon above the threshold areal density (much of which is of higher areal density than
the system penetration), which contains a shell of plutonium
and a depleted uranium tamper, as a contiguous region of $r_\text{eff}=9.6$~cm and $A_\text{eff}=272$~cm$^2$ surrounded
by the less dense chemical explosive material.}
\label{fig:si_fpu}
\end{figure}

Using the model for the measurement uncertainties described above, each of the test objects listed
in Table~\ref{tab:objs} was modeled to create simulated images of the objects with the same resolution, precision,
etc.\ as would be expected if they were imaged using the Eagle R60\textsuperscript{\textregistered} system.
First, each object was mapped to the $5\times5$~mm pixel resolution of the scanner and the material composition
of the object along the projection of each pixel was determined.  The system model (Equations~\ref{eq:tran} and \ref{eq:sysresp}) was
then applied to determine the expected transmission $T$ for each pixel, which was then converted to the
steel equivalent thickness $S$.  To determine the expected reconstructed size of each object (\reff and $A_\text{eff}$) given the pixel-to-pixel uncertainty
derived from the data each image was simulated $\mathcal{O}(1000)$ times, applying normally distributed random
fluctuations to each pixel according to the extracted standard deviation as a function of cargo thickness (\sfref{fig:si_unc}).  Additionally, since
some source variation was still present in the images after preprocessing \cite{rogers}, this effect was also simulated for the test object images.  By examining
the image set, it was determined that the remaining vertical striping artifact from the source variation amounted to an approximately 4\% reduction in the reconstructed
steel equivalent thickness of a given pixel in such a stripe.  Since this source variation could potentially cause dense regions to be split by the vertical striping,
these were conservatively simulated in test object images by applying a 10\% reduction in the reconstructed thickness in the stripes and modeling the stripes
with the maximum width observed in data.  Additionally, the thickness thresholds chosen for the analysis were deliberately chosen to be less than the expected
thicknesses of each object to place the thresholds well below this effect.  Each of these
simulated images was then processed using the same object finding algorithm as was applied to the cargo container images to compute the
expected \reff and \Aeff values for the objects.  For the objects of interest, it was found that the pixel-to-pixel variations led to
variations of at most a few percent (except for the smallest test objects, the 100 cm$^3$ sample and WGPu critical mass, which varied by
approximately $\pm$0.1 cm in radius and $\pm$1~cm$^2$ in area due to statistical fluctuation), and thus the mean values of \reff and \Aeff are quoted in Table~\ref{tab:objs}.  Figure~\ref{fig:si_fpu}
shows the application of this procedure for the Fetter, et al.\ plutonium weapon model \cite{fetter} at an object detection threshold of 25.6~cm-steel equivalent.
The algorithm identifies the dense core of the weapon (the plutonium plus the depleted uranium tamper) with the expected effective radius and area.

\section{Results for 40 Foot Containers}
\label{sec:si_40}

\begin{figure}[ht]
\centering
\includegraphics[width=\columnwidth]{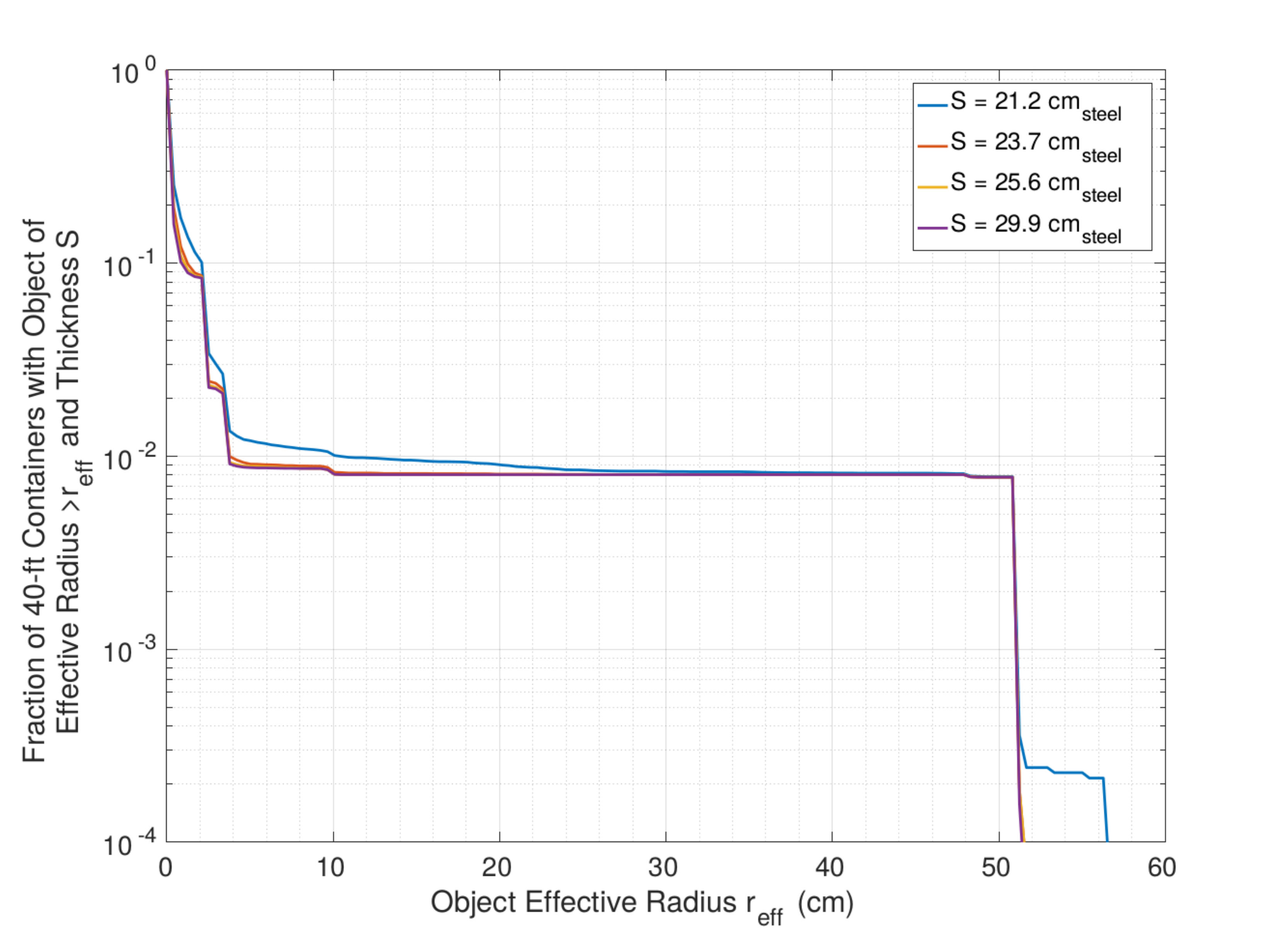}
\caption{Fraction of 40~foot container images containing a contiguous region with effective areal density ${\geq}S$ of effective radius ${\geq}r_\text{eff}$, for
several values of $S$.  See Figure~\ref{fig:r20dist} in the main article for the equivalent figure for 20~foot containers.}
\label{fig:r40dist}
\end{figure}

\begin{figure}[ht]
\centering
\includegraphics[width=\columnwidth]{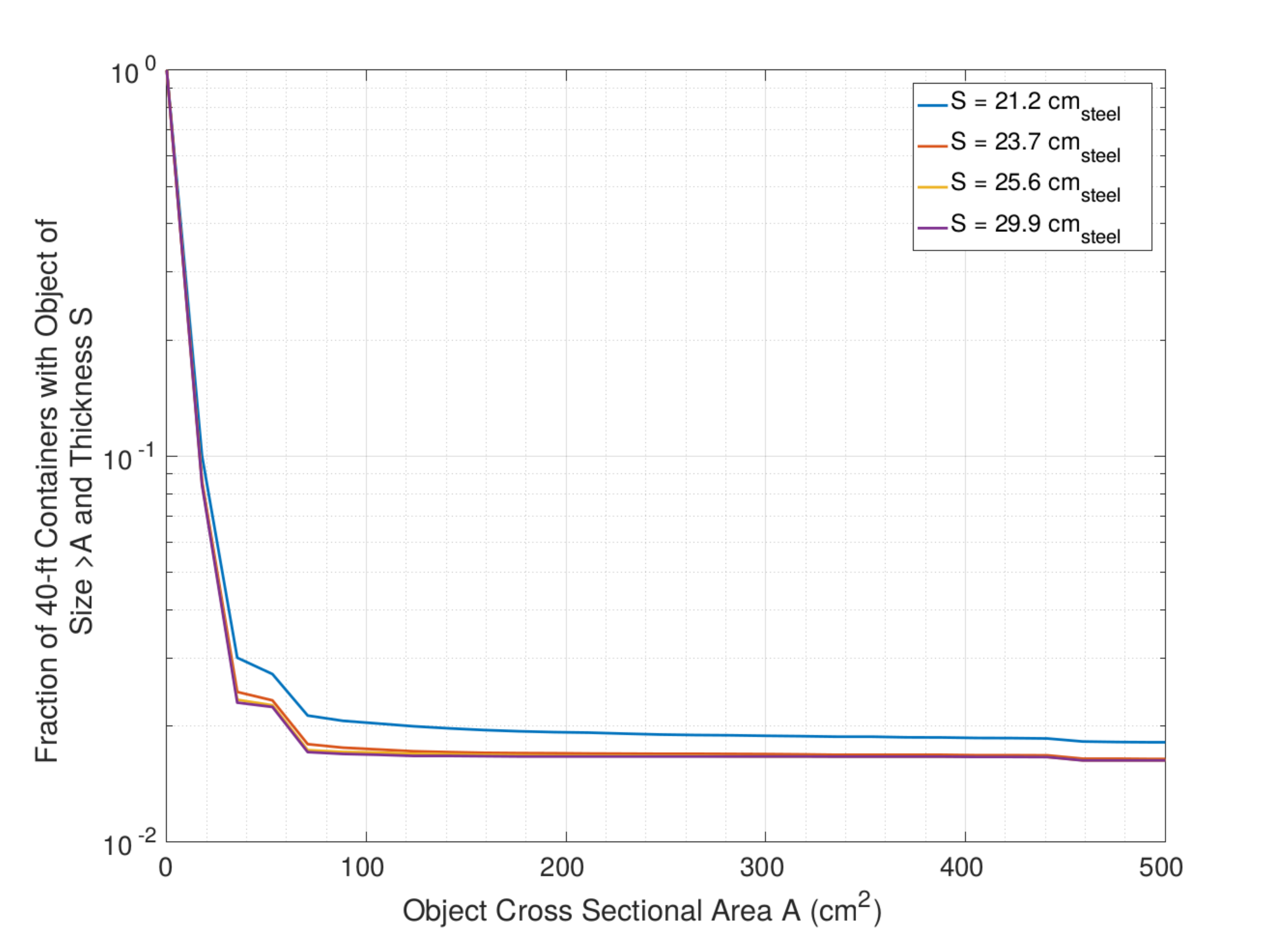}
\caption{Fraction of 40~foot container images containing a contiguous region with effective areal density ${\geq}S$ of cross sectional area
${\geq}A_\text{eff}$, for
several values of $S$.  See Figure~\ref{fig:r20dist} in the main article for the equivalent figure for 20~foot containers.}
\label{fig:a40dist}
\end{figure}

\begin{figure}[ht]
\centering
\includegraphics[width=\columnwidth]{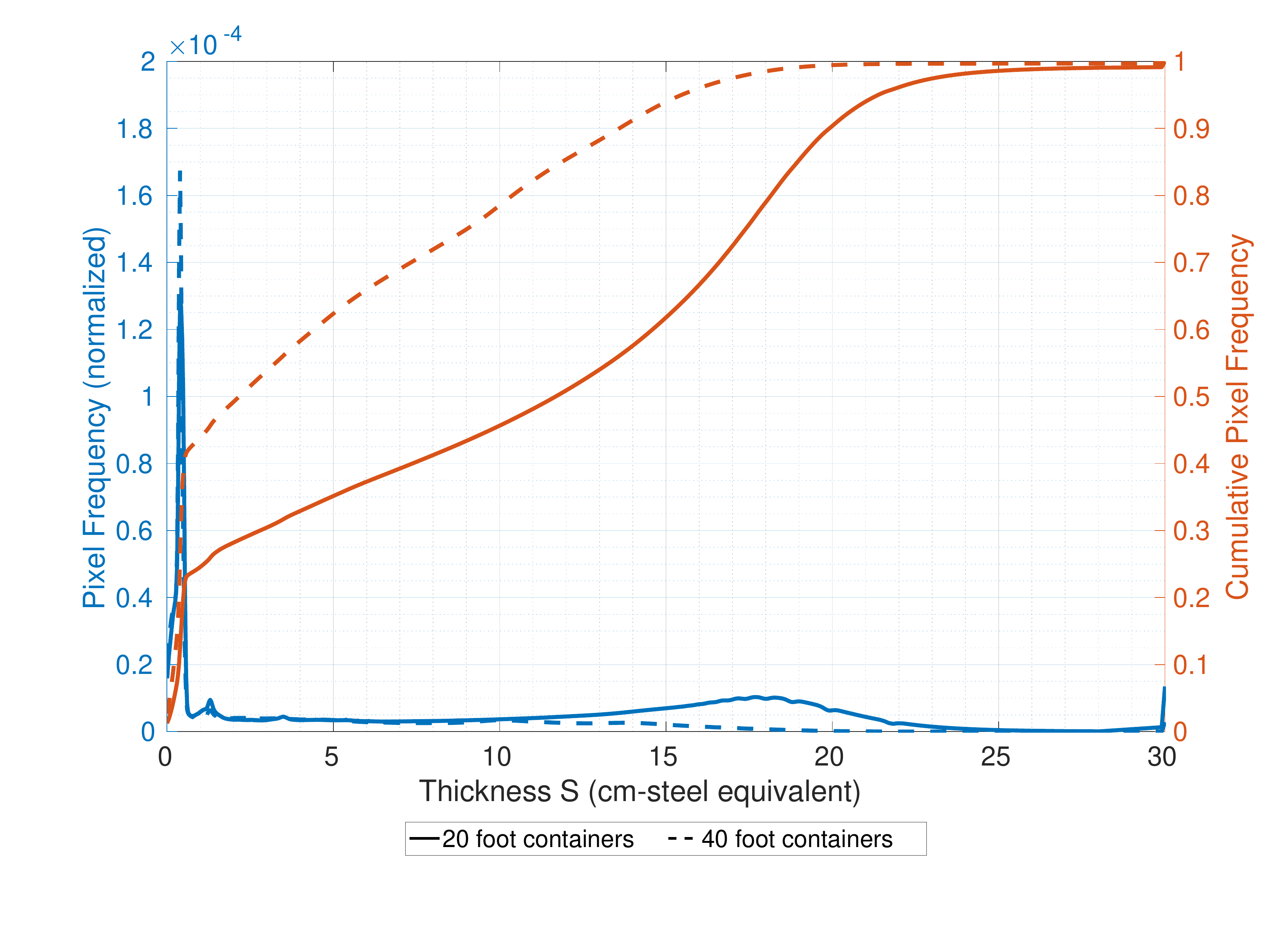}
\caption{Distributions of the effective areal density by pixel of the cargo,
separated by 20 and 40~foot containers (left axis) with the corresponding
cumulative distributions (right axis) in units of centimeters of steel
equivalent. This figure presents the same data as Figure~\ref{fig:pixdist} without truncation of the y-axis.
For these distributions, the portions of the container images including the container roofs were excluded.}
\label{fig:si_upix}
\end{figure}

The plots presented in this section (Figures~\ref{fig:r40dist} and \ref{fig:a40dist}) present the
equivalent results for 40~foot containers as are presented for 20~foot containers in Figures~\ref{fig:r40dist} and \ref{fig:a40dist}
in the main text.  As noted in Footnote~\ref{fn:40}, 40~foot containers contain
significantly fewer large, dense objects than 20 foot containers and thus the techniques discussed for identifying
threats in the main text will perform at least as well (and in many cases significantly better) when applied to
streams of 40~foot containers compared to 20~foot containers.  This result follows intuitively from the fact
that the weight limit for 40~foot containers is only slightly larger than the 20~foot
container limit in most jurisdictions \cite{im}, and thus 40-foot containers often contain spatially
large/bulky, but not dense, shipments such as large pieces of equipment.
As for the 20~foot container results shown in the main article, the data shown in this section focus
on effective object thicknesses $S>20$ cm-steel equivalent
due to the relevance of this density range to the search for objects resembling nuclear/radiological
threats. 

\section{Additional Results from the Radiographic Image Set}
\label{sec:si_add}

This section presents several other results from the image set that do not necessarily directly
pertain to the analysis presented in the main article, but may be of interest to those
working in the fields of cargo security and logistics.  In particular, these results
present information that may be useful for providing information for priors in Bayesian
analyses of cargo inspection data \cite{pmlr-v80-zheng18b} and in giving context to
previous analyses of radiography for nuclear threat detection \cite{Gaukler2011}.  The results
in this section include calculations of the average cargo areal density as a function of
position in containers (Figures~\ref{fig:hmean} and \ref{fig:vmean}) and the  distributions of the sizes of of the largest
objects by container that correspond to the cumulative distributions presented in the main text.
Each result is described in the caption accompanying its figure, along with any notes regarding
interesting features of the data.




\begin{figure}[ht]
\centering
\includegraphics[width=\columnwidth]{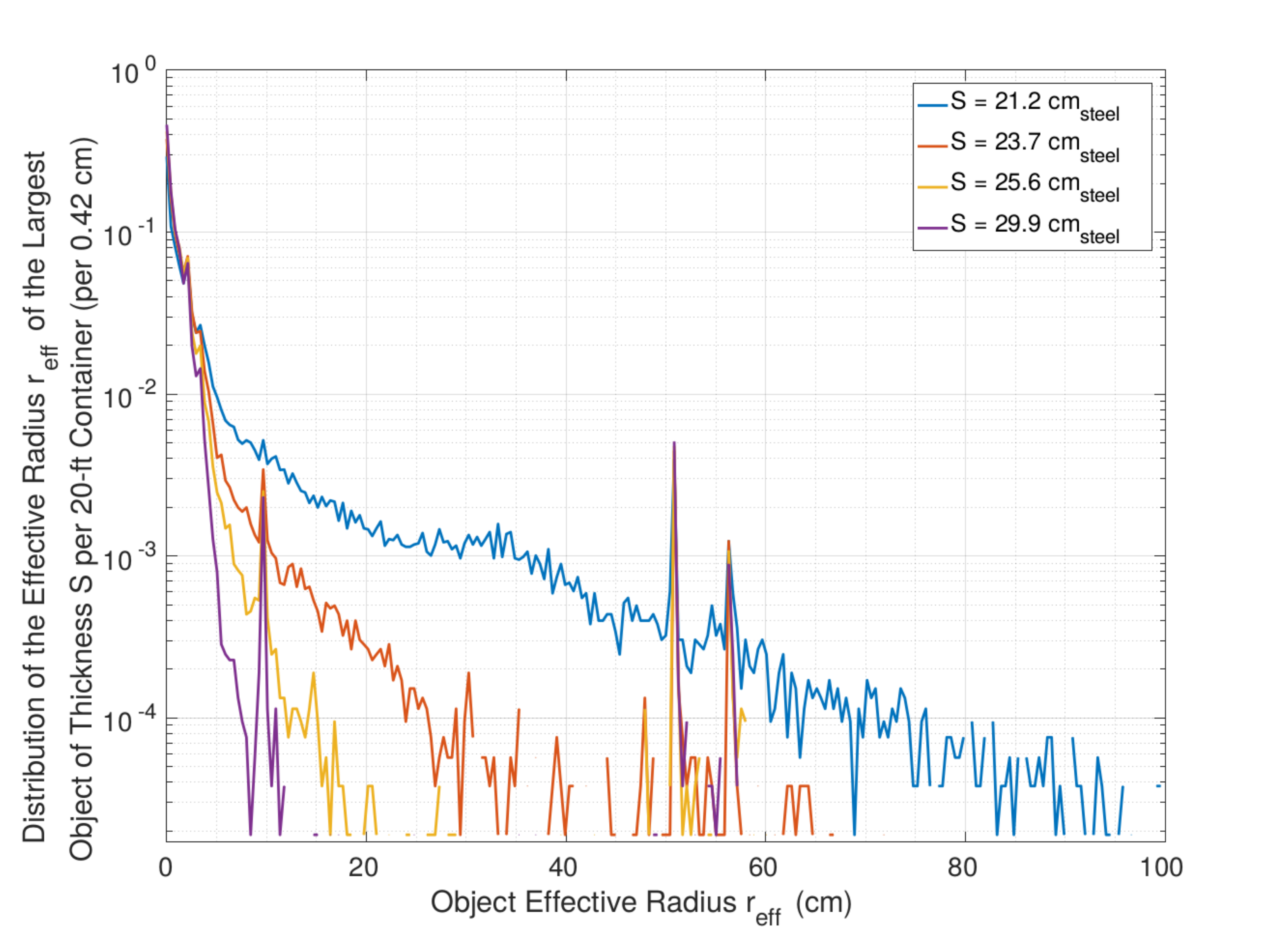}
\caption{Distribution of the largest contiguous region by \reff with effective areal density ${\geq}S$ per 20 foot container, for
several values of $S$.  Figure~\ref{fig:r20dist} represents unity minus the cumulative distribution of this probability distribution.}
\label{fig:r20p}
\end{figure}

\begin{figure}[ht]
\centering
\includegraphics[width=\columnwidth]{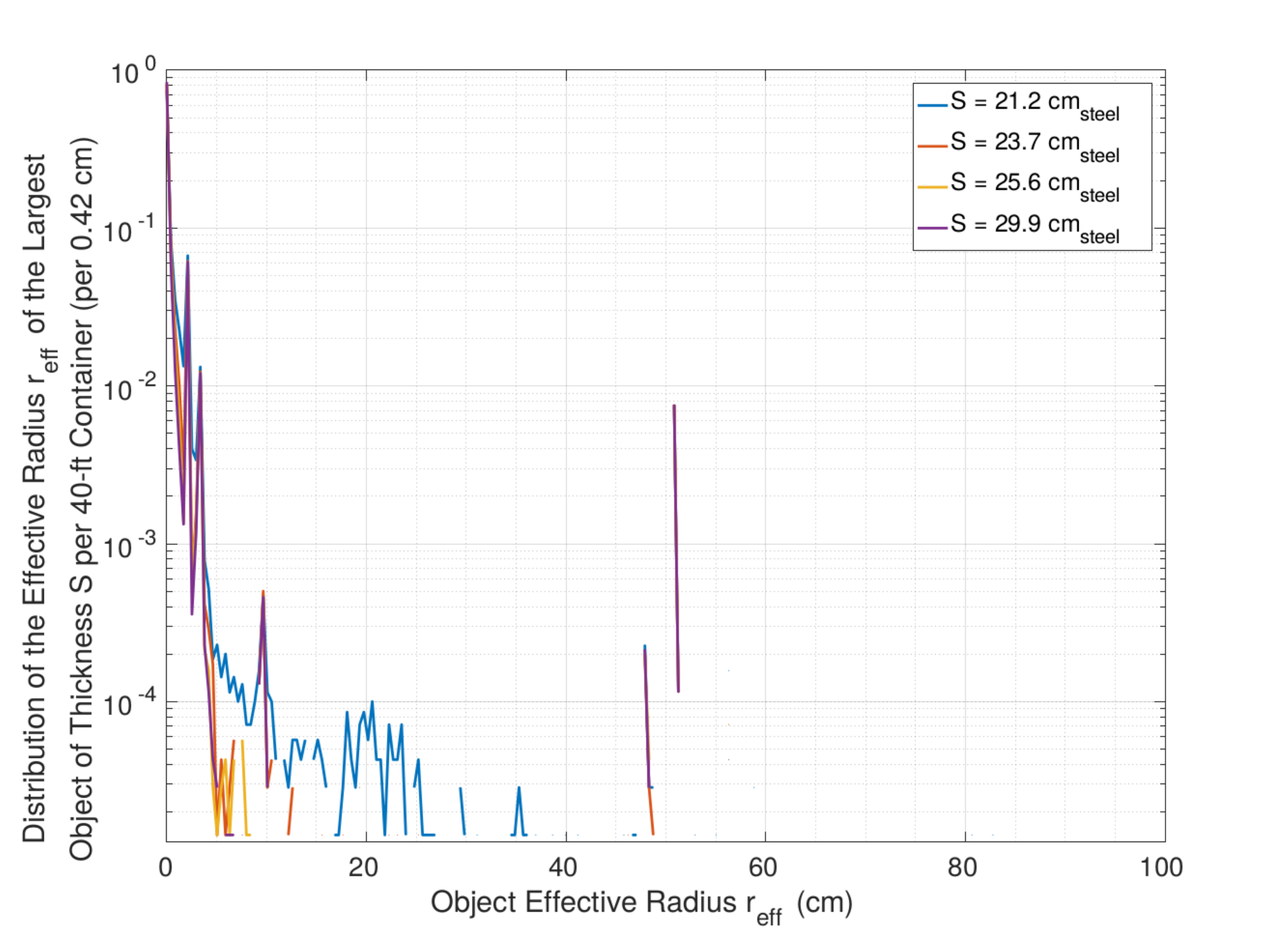}
\caption{Distribution of the largest contiguous region by \reff with effective areal density ${\geq}S$ per 40 foot container, for
several values of $S$.  \sfref{fig:r40dist} represents unity minus the cumulative distribution of this probability distribution.}
\label{fig:r40p}
\end{figure}

\begin{figure}[ht]
\centering
\includegraphics[width=\columnwidth]{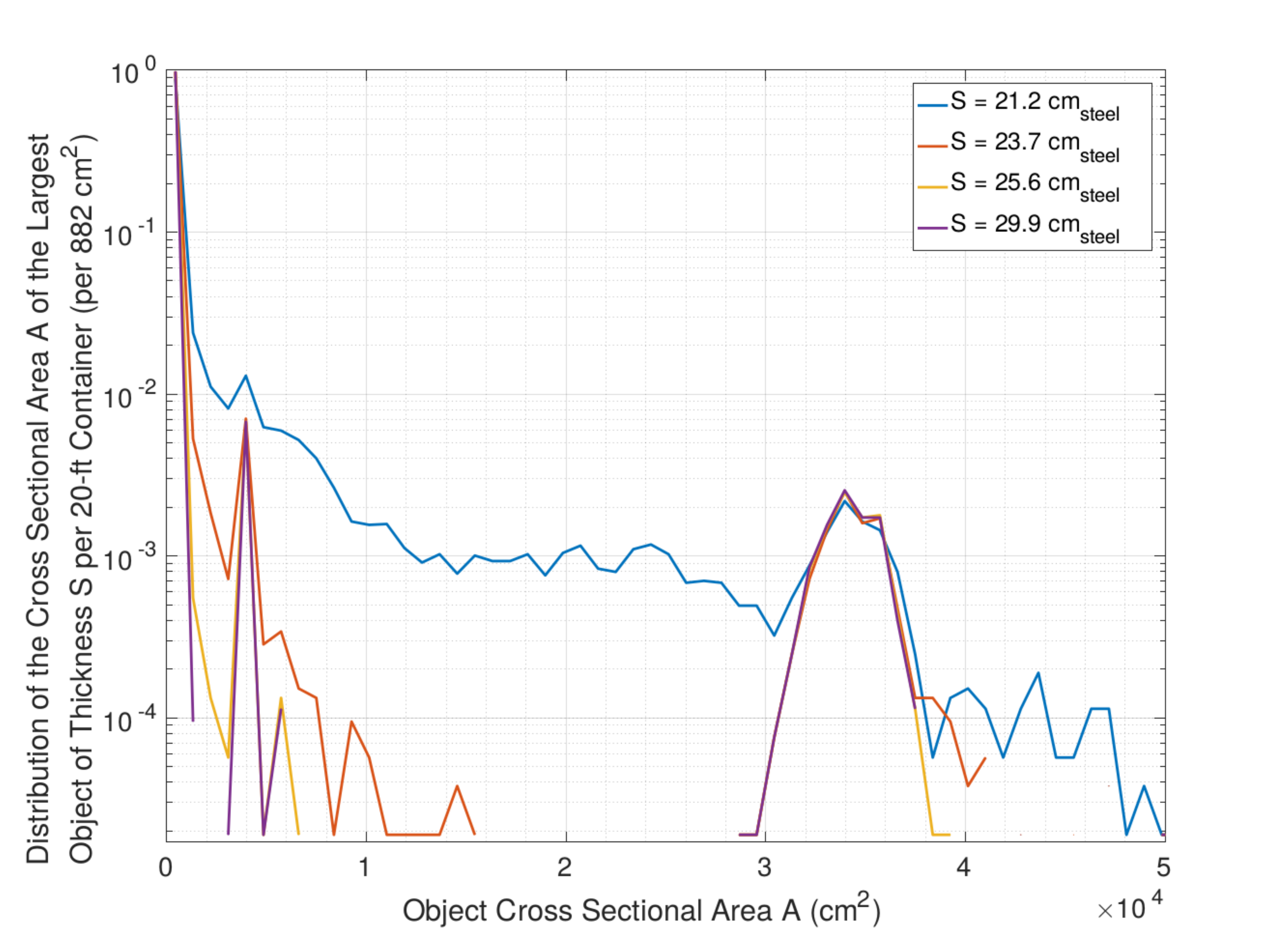}
\caption{Distribution of the largest contiguous region by \Aeff with effective areal density ${\geq}S$ per 20 foot container, for
several values of $S$.  Figure~\ref{fig:a20dist} represents unity minus the cumulative distribution of this probability distribution.}
\label{fig:a20p}
\end{figure}

\begin{figure}[ht]
\centering
\includegraphics[width=\columnwidth]{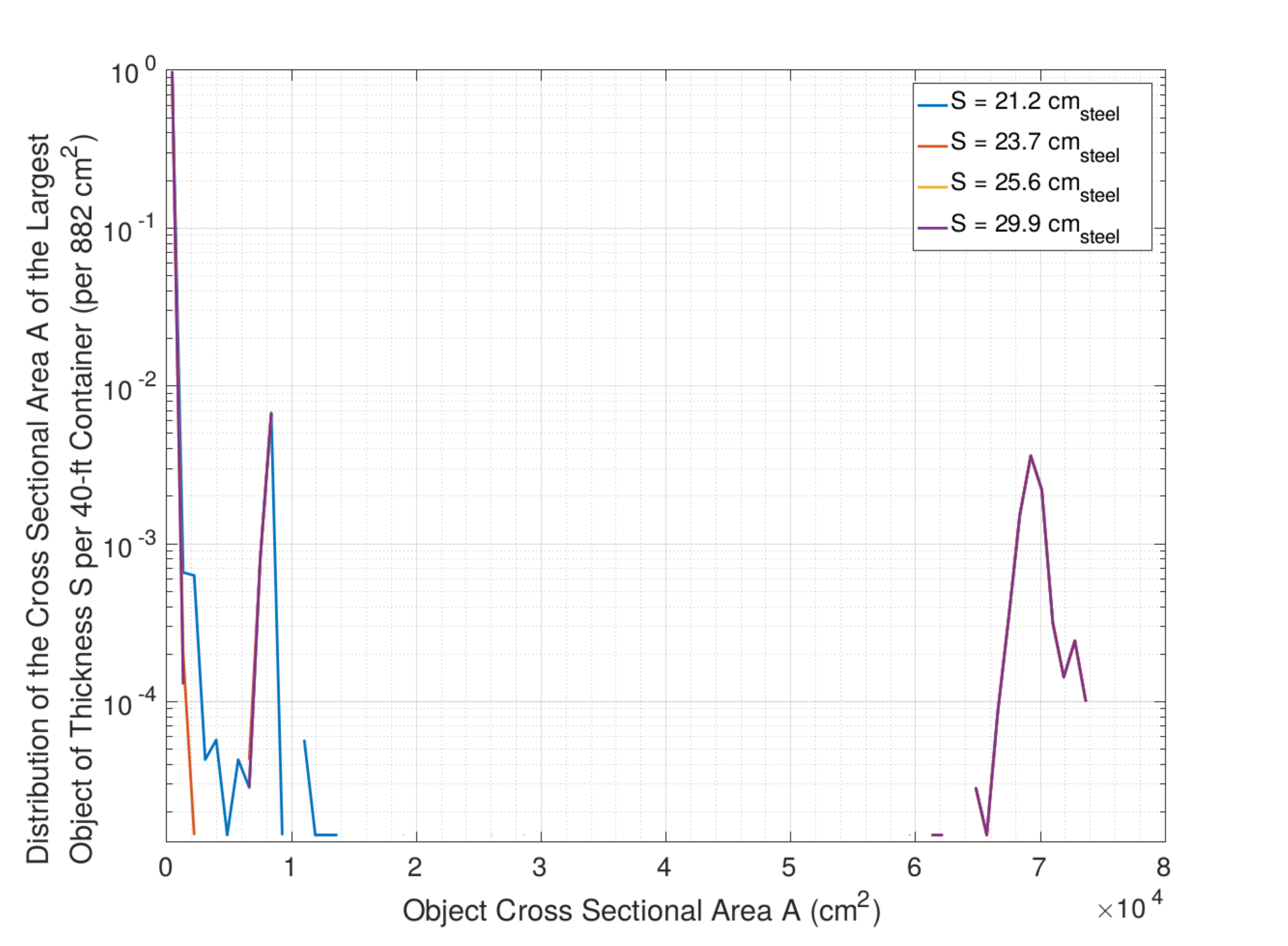}
\caption{Distribution of the largest contiguous region by \Aeff with effective areal density ${\geq}S$ per 40 foot container, for
several values of $S$.  \sfref{fig:a40dist} represents unity minus the cumulative distribution of this probability distribution.}
\label{fig:a40p}
\end{figure}

\begin{figure}[ht]
\centering
\includegraphics[width=\columnwidth]{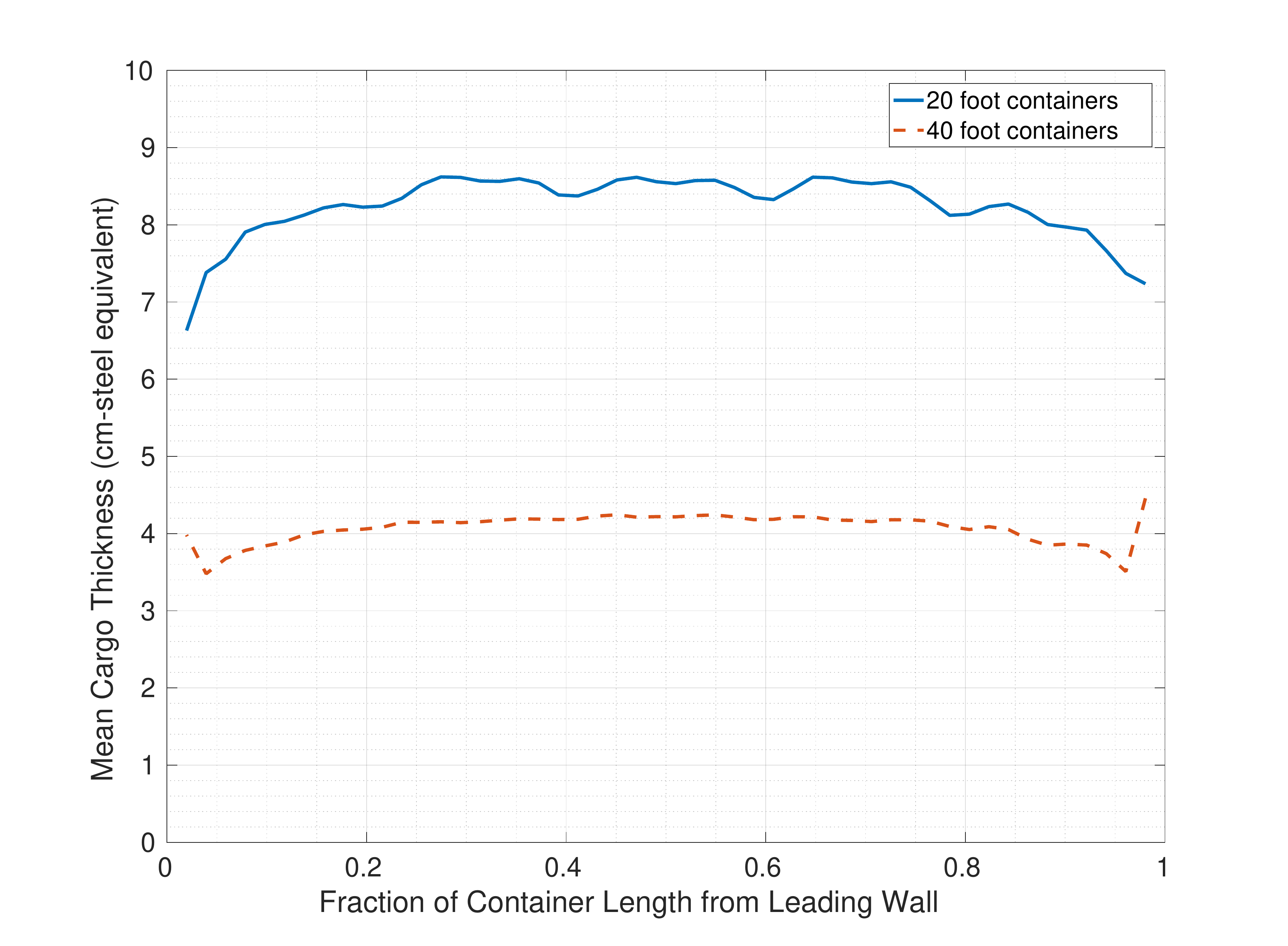}
\caption{Mean cargo thickness in centimeters of steel equivalent as a function
of position along the long axis of the containers (measured as a fraction of the distance
along the container length from the leading wall) for the 20~foot and 40~foot image sets.
Note the structure in the 20~foot data due to the common use of 48''$\times$48''$\times$48'' palettes in this
cargo stream, while the uniformity of the 40~foot container data indicates predominantly non-palletized cargo.
In both cases, there is a slight bias towards higher densities in the central region of the containers.}
\label{fig:hmean}
\end{figure}

\begin{figure}[ht]
\centering
\includegraphics[width=\columnwidth]{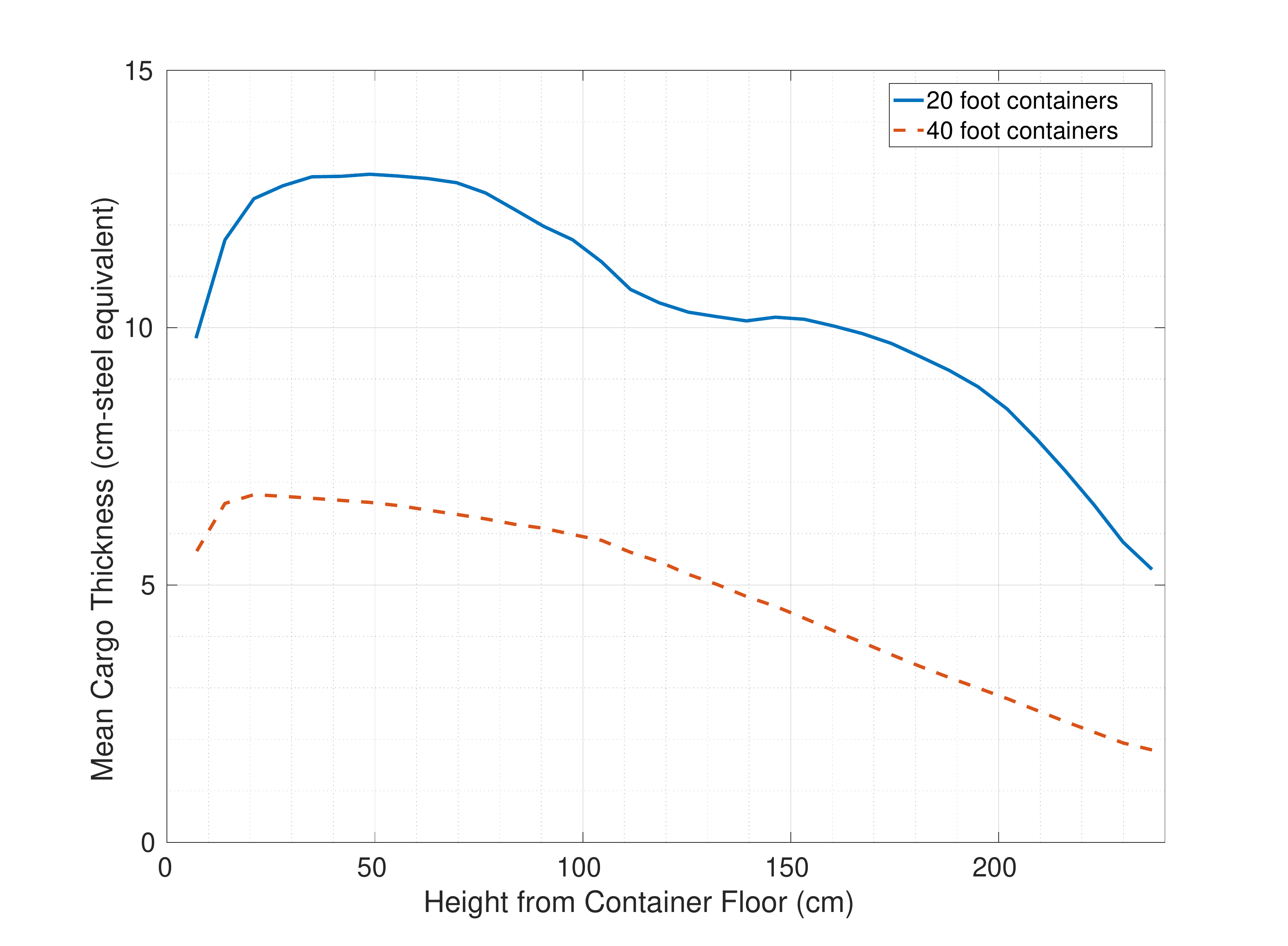}
\caption{Mean cargo thickness in centimeters of steel equivalent as a function
of height in the containers for the 20~foot and 40~foot image sets.  Both sets indicate the reasonable trend of decreasing
cargo density with increasing height in the container. Again, the 20~foot container
data indicates stacks of two 48''$\times$48''$\times$48'' palettes, while the 40~foot data shows no significant signs of
palletized cargo.  In each case, cargo is denser on average near the floors of containers as would be intuitively expected.}
\label{fig:vmean}
\end{figure}



\end{document}